\documentclass[aps,prb,superscriptaddress,sd,reprint]{revtex4-1}
\usepackage{graphicx}
\usepackage{dcolumn}
\usepackage{bm}
\usepackage[colorlinks,linkcolor=blue,citecolor=blue,urlcolor=blue]{hyperref}
\usepackage{amsmath,amsthm,amssymb}
\usepackage{amsfonts}
\usepackage{icomma}
\usepackage{epstopdf}
\usepackage{placeins}
\usepackage{color}
\usepackage[euler]{textgreek}
\usepackage[separate-uncertainty = true, exponent-product = \cdot]{siunitx}
\DeclareSIUnit\torr{Torr}

\makeatletter
\let\cat@comma@active\@empty
\makeatother

\newcommand{\GdFeCo}{Gd$_{21}($Fe$_{90}$Co$_{10})_{79}$ }

\begin{document}

\title{Thickness Dependence of Spin-Orbit Torques in Ferrimagnetic GdFeCo Alloys}

\author{Niklas Roschewsky}
\email{roschewsky@berkeley.edu}
\affiliation{Department of Physics, University of California, Berkeley, California 94720, USA}

\author{Charles-Henri Lambert}
\affiliation{Department of Electrical Engineering and Computer Sciences, University of California, Berkeley, California 94720, USA}


\author{Sayeef Salahuddin}
\email{sayeef@berkeley.edu}
\affiliation{Department of Electrical Engineering and Computer Sciences, University of California, Berkeley, California 94720, USA}
\affiliation{Materials Science Division, Lawrence Berkeley National Laboratory, Berkeley, California 94720, USA}

\date{\today}

\begin{abstract}
So far, studies of spin-orbit torques (SOT) in ferromagnets with perpendicular magnetic anisotropy (PMA) have been restricted to ultra thin samples, while a systematic study of it's thickness dependence is still lacking in literature. In this article we discuss the thickness dependence of SOT in \GdFeCo samples with bulk PMA. We show that the effective SOT fields are decreasing inversely as a function of thickness while the spin-Hall angle stays constant, as expected from angular momentum conservation. Further we show that even $\SI{30}{\nano\meter}$ thick \GdFeCo samples can be switched with SOT. This has important technological implications as the switching efficiency does not depend on the thickness. Finally, we investigate the composition dependence of SOT in $\SI{30}{\nano\meter}$ thick GdFeCo samples and find that the spin torque effective field diverges at the magnetization compensation point.
\end{abstract}

\maketitle

\section{Introduction}
Commercial magnetic random access memory (MRAM) devices have been available on the market since 2006~\cite{Apalkov2016}. Most of the MRAM devices sold today are either based on the toggle switching mechanism~\cite{Engel2005} or spin transfer torque (STT)~\cite{Rizzo2013}. However, these technologies have significant drawbacks. While toggle switching is inherently unscalable, STT-MRAM faces reliability issues due to large currents through the tunnel barrier as well as slow switching times due to the precessional switching.

These challenges could be overcome with the introduction of 3-terminal memory devices utilizing spin-orbit torque (SOT) as a mechanism to write information and tunnel magneto-resistance as a mechanism to read~\cite{Lee2016}. Here SOT refers to the generation of a non-equilibrium spin-accumulation at a heavy metal (HM)/ferromagnet (FM) interface, either due to the bulk spin-Hall effect in the HM~\cite{Liu2012a,Liu2012b} or the Rashba-Edelstein effect at the HM/FM interface~\cite{Miron2011}. This non-equilibrium spin-accumulation diffuses into the ferromagnet~\cite{Hane2013}, where it can reverse the magnetic order via the spin-transfer torque mechanism~\cite{Berger1996,Slonczewski1996}.

These 3-terminal memory devices require magnetic materials which exhibit perpendicular magnetic anisotropy (PMA). In most devices today, transition metal magnets, such as Co or Fe, with MgO capping layers are used~\cite{Ikeda2010}. Here, the PMA originates from hybridization effects at the ferromagnet/oxide interface\cite{Yang2011}. Due to the inherently interfacial nature of this anisotropy, the ferromagnetic films must be grown very thin (on the order of a \SI{1}{\nano\meter}) and clean interfaces with the right oxygen stoichiometry are important.

More recently, SOT have been investigated in HM/ferrimagnet structures, where the ferrimagnets are transitions metal (TM)-rare earth (RE) alloys~\cite{Zhao2015,Ueda2016,Roschewsky2016,Finley2016}. These materials exhibit bulk PMA. This has several technological advantages. For instance, interfaces can be altered without changing the magnetic properties drastically. Further, the magnetic properties in ferrimagnetic alloys can be tuned by changing the film composition. Finally, the ferrimagnetic film thickness can be used as a new design parameter for SOT devices. Due to the bulk nature of the PMA, ferrimagnetic films can be grown much thicker than ferromagnet/oxide structures discussed above.

So far however, a detailed study of the thickness dependence of SOT is still lacking in literature. While this has been attempted before in magnets with in-plane anisotropy~\cite{Fan2014}, studies in magnets with PMA  remain challenging, either due to changes in anisotropy energy with thickness~\cite{Conte2016} or changes in the crystalline structure with thickness~\cite{Lee2014}. Therefore, all studies of SOT in films with PMA to date are limited to ultra thin magnetic films of just a few nanometers thickness.

\begin{figure}[b]
\begin{center}
\includegraphics[width=\columnwidth]{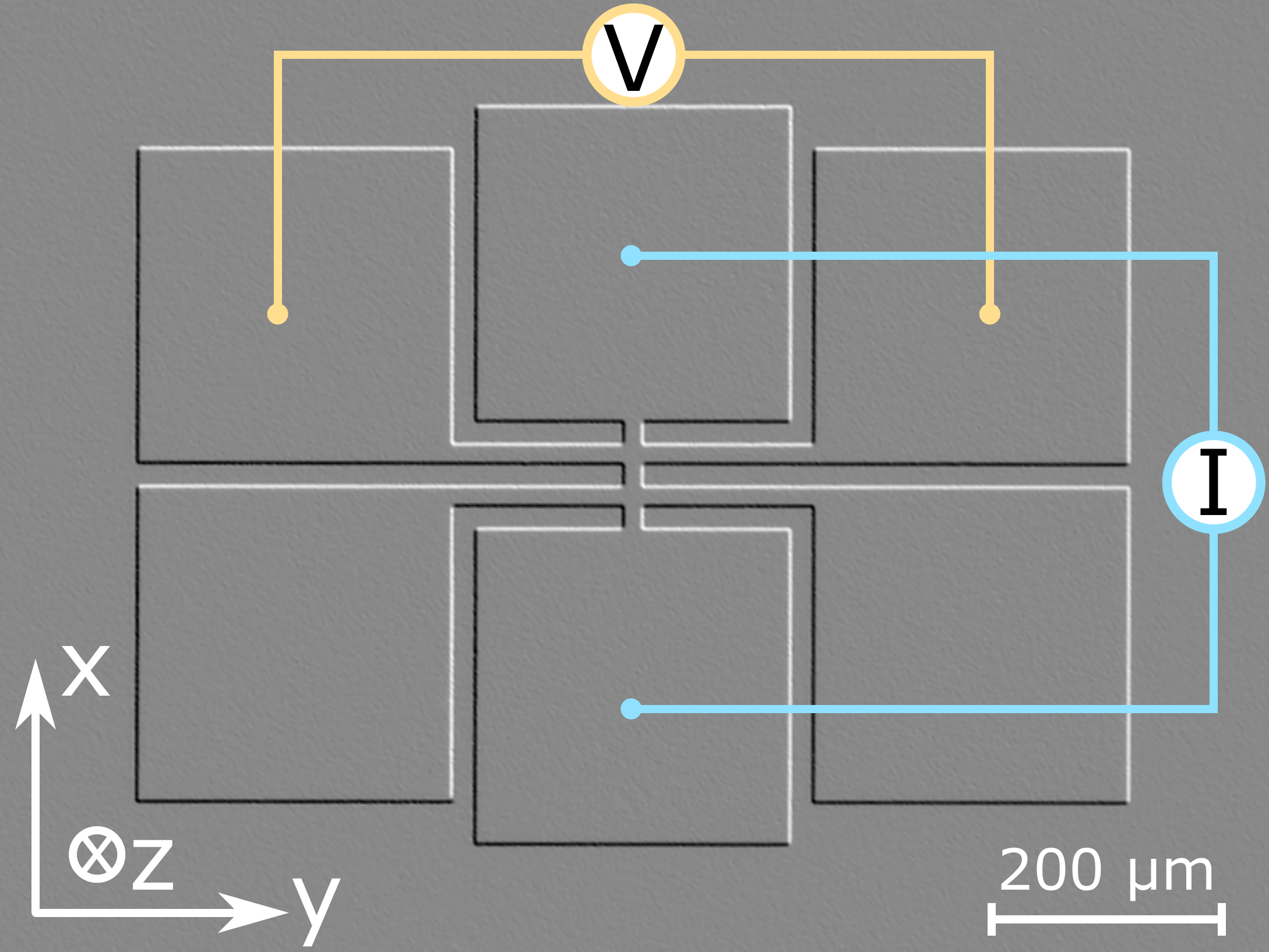}
\caption{\label{fig:Hallbar} Microscope image of a Hall bar device. Ta$(5)$/Gd$_{21}($Fe$_{90}$Co$_{10})_{79}(t)$/Pt$(5)$ (thickness in $\si{\nano\meter}$) structures are investigated. A current is applied along the $x$-direction and the Hall voltage is detected along the $y$-direction. The width of the Hall bar is $\SI{20}{\micro\meter}$.}
\end{center}
\end{figure}
In this study we investigate the thickness dependence of SOT in ferrimagnetic \GdFeCo films. We vary the thicknesses from \SI{10}{\nano\meter} up to \SI{30}{\nano\meter} and show that all films can be switched with current pulses through adjacent HM layers. To measure the effective magnetic field due to SOT, we perform harmonic Hall measurements. The effective field is found to be inversely proportional to the ferrimagnet film thickness and saturation magnetization. In addition DC transport measurements are performed. It is found that the critical switching current $j_c$, as well as the thermal stability $\Delta$ increase linearly with thickness, so that the switching efficiency $j_c/\Delta$ is constant. This result has important implications for memory technology, as the film thickness can be tuned in order to archive desired values for the critical current and thermal stability.

\begin{figure}[t]
\begin{center}
%
%
\begingroup
  \makeatletter
  \providecommand\color[2][]{%
    \GenericError{(gnuplot) \space\space\space\@spaces}{%
      Package color not loaded in conjunction with
      terminal option `colourtext'%
    }{See the gnuplot documentation for explanation.%
    }{Either use 'blacktext' in gnuplot or load the package
      color.sty in LaTeX.}%
    \renewcommand\color[2][]{}%
  }%
  \providecommand\includegraphics[2][]{%
    \GenericError{(gnuplot) \space\space\space\@spaces}{%
      Package graphicx or graphics not loaded%
    }{See the gnuplot documentation for explanation.%
    }{The gnuplot epslatex terminal needs graphicx.sty or graphics.sty.}%
    \renewcommand\includegraphics[2][]{}%
  }%
  \providecommand\rotatebox[2]{#2}%
  \@ifundefined{ifGPcolor}{%
    \newif\ifGPcolor
    \GPcolortrue
  }{}%
  \@ifundefined{ifGPblacktext}{%
    \newif\ifGPblacktext
    \GPblacktextfalse
  }{}%
  \let\gplgaddtomacro\g@addto@macro
  \gdef\gplbacktext{}%
  \gdef\gplfronttext{}%
  \makeatother
  \ifGPblacktext
    \def\colorrgb#1{}%
    \def\colorgray#1{}%
  \else
    \ifGPcolor
      \def\colorrgb#1{\color[rgb]{#1}}%
      \def\colorgray#1{\color[gray]{#1}}%
      \expandafter\def\csname LTw\endcsname{\color{white}}%
      \expandafter\def\csname LTb\endcsname{\color{black}}%
      \expandafter\def\csname LTa\endcsname{\color{black}}%
      \expandafter\def\csname LT0\endcsname{\color[rgb]{1,0,0}}%
      \expandafter\def\csname LT1\endcsname{\color[rgb]{0,1,0}}%
      \expandafter\def\csname LT2\endcsname{\color[rgb]{0,0,1}}%
      \expandafter\def\csname LT3\endcsname{\color[rgb]{1,0,1}}%
      \expandafter\def\csname LT4\endcsname{\color[rgb]{0,1,1}}%
      \expandafter\def\csname LT5\endcsname{\color[rgb]{1,1,0}}%
      \expandafter\def\csname LT6\endcsname{\color[rgb]{0,0,0}}%
      \expandafter\def\csname LT7\endcsname{\color[rgb]{1,0.3,0}}%
      \expandafter\def\csname LT8\endcsname{\color[rgb]{0.5,0.5,0.5}}%
    \else
      \def\colorrgb#1{\color{black}}%
      \def\colorgray#1{\color[gray]{#1}}%
      \expandafter\def\csname LTw\endcsname{\color{white}}%
      \expandafter\def\csname LTb\endcsname{\color{black}}%
      \expandafter\def\csname LTa\endcsname{\color{black}}%
      \expandafter\def\csname LT0\endcsname{\color{black}}%
      \expandafter\def\csname LT1\endcsname{\color{black}}%
      \expandafter\def\csname LT2\endcsname{\color{black}}%
      \expandafter\def\csname LT3\endcsname{\color{black}}%
      \expandafter\def\csname LT4\endcsname{\color{black}}%
      \expandafter\def\csname LT5\endcsname{\color{black}}%
      \expandafter\def\csname LT6\endcsname{\color{black}}%
      \expandafter\def\csname LT7\endcsname{\color{black}}%
      \expandafter\def\csname LT8\endcsname{\color{black}}%
    \fi
  \fi
    \setlength{\unitlength}{0.0500bp}%
    \ifx\gptboxheight\undefined%
      \newlength{\gptboxheight}%
      \newlength{\gptboxwidth}%
      \newsavebox{\gptboxtext}%
    \fi%
    \setlength{\fboxrule}{0.5pt}%
    \setlength{\fboxsep}{1pt}%
\begin{picture}(4874.00,3400.00)%
    \gplgaddtomacro\gplbacktext{%
      \csname LTb\endcsname%
      \put(682,704){\makebox(0,0)[r]{\strut{}$4$}}%
      \put(682,1312){\makebox(0,0)[r]{\strut{}$8$}}%
      \put(682,1920){\makebox(0,0)[r]{\strut{}$12$}}%
      \put(682,2527){\makebox(0,0)[r]{\strut{}$16$}}%
      \put(682,3135){\makebox(0,0)[r]{\strut{}$20$}}%
      \put(814,484){\makebox(0,0){\strut{}$5$}}%
      \put(1300,484){\makebox(0,0){\strut{}$10$}}%
      \put(1786,484){\makebox(0,0){\strut{}$15$}}%
      \put(2272,484){\makebox(0,0){\strut{}$20$}}%
      \put(2757,484){\makebox(0,0){\strut{}$25$}}%
      \put(3243,484){\makebox(0,0){\strut{}$30$}}%
      \put(3729,484){\makebox(0,0){\strut{}$35$}}%
      \colorrgb{0.02,0.73,0.42}%
      \put(3861,704){\makebox(0,0)[l]{\strut{}$0$}}%
      \colorrgb{0.02,0.73,0.42}%
      \put(3861,1312){\makebox(0,0)[l]{\strut{}$20$}}%
      \colorrgb{0.02,0.73,0.42}%
      \put(3861,1920){\makebox(0,0)[l]{\strut{}$40$}}%
      \colorrgb{0.02,0.73,0.42}%
      \put(3861,2527){\makebox(0,0)[l]{\strut{}$60$}}%
      \colorrgb{0.02,0.73,0.42}%
      \put(3861,3135){\makebox(0,0)[l]{\strut{}$80$}}%
      \colorrgb{0.87,0.09,0.12}%
      \put(1397,1069){\rotatebox{15}{\makebox(0,0)[l]{\strut{}$10\cdot\rho_\text{PHE}$}}}%
      \colorrgb{0.00,0.38,0.68}%
      \put(1494,1464){\rotatebox{36}{\makebox(0,0)[l]{\strut{}$\rho_\text{AHE}$}}}%
      \colorrgb{0.02,0.73,0.42}%
      \put(1300,2497){\makebox(0,0)[l]{\strut{}$M_\text s$}}%
      \csname LTb\endcsname%
      \put(1980,2983){\makebox(0,0)[l]{\strut{}$\text{Gd}_{21}(\text{Fe}_{90}\text{Co}_{10})_{79}(t)$}}%
    }%
    \gplgaddtomacro\gplfronttext{%
      \csname LTb\endcsname%
      \put(176,1919){\rotatebox{-270}{\makebox(0,0){\strut{}$\rho$ (\textmu\textOmega$\cdot$cm)}}}%
      \colorrgb{0.02,0.73,0.42}%
      \put(4366,1919){\rotatebox{-270}{\makebox(0,0){\strut{}$M_s$ (kA/m)}}}%
      \csname LTb\endcsname%
      \put(2271,154){\makebox(0,0){\strut{}$t$ (nm)}}%
    }%
    \gplbacktext
    \put(0,0){\includegraphics{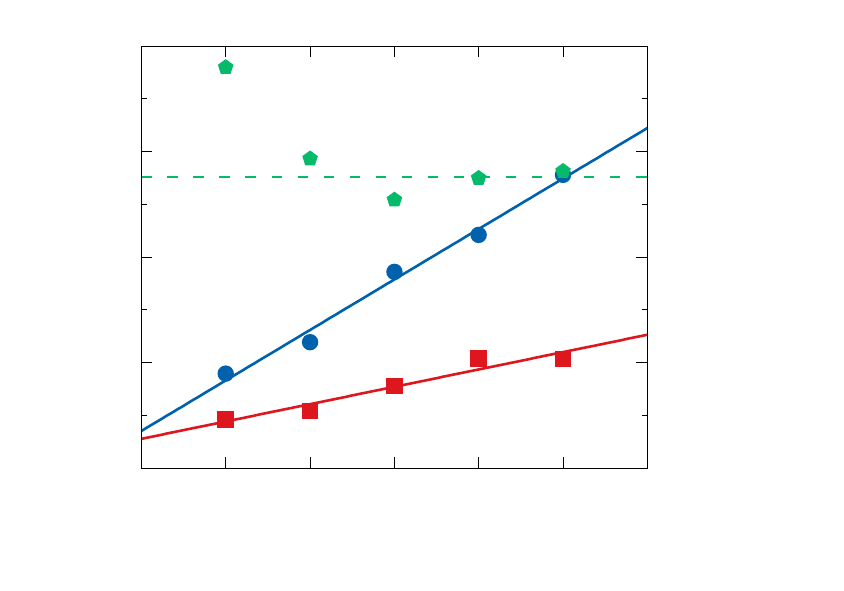}}%
    \gplfronttext
  \end{picture}%
\endgroup
\caption{\label{fig:Resistivity} The AHE resistivity $\rho_\text{AHE}$ (blue, left axis) as well as the PHE resistivity $\rho_\text{PHE}$ (red, left axis) depend linearly on the \GdFeCo thickness. The saturation magnetization was measured with VSM magnetometry and is plotted in green on the right axis. $M_\text S$ is constant for samples with thickness $t>\SI{10}{\nano\meter}$. At $t=\SI{10}{\nano\meter}$, the magnetization is enhanced.}
\end{center}
\end{figure}
\section{Sample Preparation}
To investigate the thickness dependence of SOT in ferrimagnets we deposited a series of Ta$(5)$/Gd$_{21}($Fe$_{90}$Co$_{10})_{79}(t)$/Pt$(5)$ films (thickness in \si{\nano\meter}) on thermally oxidized silicon substrates by RF magnetron sputtering. The thickness $t$ was varied from $\SI{10}{\nano\meter}$ to $\SI{30}{\nano\meter}$ in steps of $\SI{5}{\nano\meter}$. The base pressure during deposition was below \SI{1e-8}{\torr}. After deposition, Hall bar mesa structures were defined by optical lithography and Ar-ion milling. Figure~\ref{fig:Hallbar} shows an optical image of a typical Hall bar device and the measurement geometry. The width of the Hall bar is $\SI{20}{\micro\meter}$. A current is applied along the $x$-direction to excite SOT dynamics. The response of the magnet is detected by the AHE, measured perpendicular to the current direction, along $y$. Note that Gd$_{21}($Fe$_{90}$Co$_{10})_{79}$ is a transition metal rich alloy with $|m_\text{TM}|>|m_\text{RE}|$. The resistivity of the \GdFeCo is $\rho=\SI{1.19}{\milli\ohm\centi\meter}$.

\section{Results and Discussion}
The anomalous Hall effect (AHE) in TM-RE ferrimagnets is proportional to the out-of-plane component of the TM magnetization, while the RE magnet does not contribute significantly the the AHE~\cite{Shirakawa1976,Mimura1976}. This is because the conduction electrons in the TM are spin polarized while the RE has no spin-split conduction band. 

\begin{figure}[t]
\begin{center}
%
%
\begingroup
  \makeatletter
  \providecommand\color[2][]{%
    \GenericError{(gnuplot) \space\space\space\@spaces}{%
      Package color not loaded in conjunction with
      terminal option `colourtext'%
    }{See the gnuplot documentation for explanation.%
    }{Either use 'blacktext' in gnuplot or load the package
      color.sty in LaTeX.}%
    \renewcommand\color[2][]{}%
  }%
  \providecommand\includegraphics[2][]{%
    \GenericError{(gnuplot) \space\space\space\@spaces}{%
      Package graphicx or graphics not loaded%
    }{See the gnuplot documentation for explanation.%
    }{The gnuplot epslatex terminal needs graphicx.sty or graphics.sty.}%
    \renewcommand\includegraphics[2][]{}%
  }%
  \providecommand\rotatebox[2]{#2}%
  \@ifundefined{ifGPcolor}{%
    \newif\ifGPcolor
    \GPcolortrue
  }{}%
  \@ifundefined{ifGPblacktext}{%
    \newif\ifGPblacktext
    \GPblacktextfalse
  }{}%
  \let\gplgaddtomacro\g@addto@macro
  \gdef\gplbacktext{}%
  \gdef\gplfronttext{}%
  \makeatother
  \ifGPblacktext
    \def\colorrgb#1{}%
    \def\colorgray#1{}%
  \else
    \ifGPcolor
      \def\colorrgb#1{\color[rgb]{#1}}%
      \def\colorgray#1{\color[gray]{#1}}%
      \expandafter\def\csname LTw\endcsname{\color{white}}%
      \expandafter\def\csname LTb\endcsname{\color{black}}%
      \expandafter\def\csname LTa\endcsname{\color{black}}%
      \expandafter\def\csname LT0\endcsname{\color[rgb]{1,0,0}}%
      \expandafter\def\csname LT1\endcsname{\color[rgb]{0,1,0}}%
      \expandafter\def\csname LT2\endcsname{\color[rgb]{0,0,1}}%
      \expandafter\def\csname LT3\endcsname{\color[rgb]{1,0,1}}%
      \expandafter\def\csname LT4\endcsname{\color[rgb]{0,1,1}}%
      \expandafter\def\csname LT5\endcsname{\color[rgb]{1,1,0}}%
      \expandafter\def\csname LT6\endcsname{\color[rgb]{0,0,0}}%
      \expandafter\def\csname LT7\endcsname{\color[rgb]{1,0.3,0}}%
      \expandafter\def\csname LT8\endcsname{\color[rgb]{0.5,0.5,0.5}}%
    \else
      \def\colorrgb#1{\color{black}}%
      \def\colorgray#1{\color[gray]{#1}}%
      \expandafter\def\csname LTw\endcsname{\color{white}}%
      \expandafter\def\csname LTb\endcsname{\color{black}}%
      \expandafter\def\csname LTa\endcsname{\color{black}}%
      \expandafter\def\csname LT0\endcsname{\color{black}}%
      \expandafter\def\csname LT1\endcsname{\color{black}}%
      \expandafter\def\csname LT2\endcsname{\color{black}}%
      \expandafter\def\csname LT3\endcsname{\color{black}}%
      \expandafter\def\csname LT4\endcsname{\color{black}}%
      \expandafter\def\csname LT5\endcsname{\color{black}}%
      \expandafter\def\csname LT6\endcsname{\color{black}}%
      \expandafter\def\csname LT7\endcsname{\color{black}}%
      \expandafter\def\csname LT8\endcsname{\color{black}}%
    \fi
  \fi
    \setlength{\unitlength}{0.0500bp}%
    \ifx\gptboxheight\undefined%
      \newlength{\gptboxheight}%
      \newlength{\gptboxwidth}%
      \newsavebox{\gptboxtext}%
    \fi%
    \setlength{\fboxrule}{0.5pt}%
    \setlength{\fboxsep}{1pt}%
\begin{picture}(4874.00,5102.00)%
    \gplgaddtomacro\gplbacktext{%
      \colorrgb{0.00,0.38,0.68}%
      \put(792,3255){\makebox(0,0)[r]{\strut{}$5.2$}}%
      \colorrgb{0.00,0.38,0.68}%
      \put(792,3782){\makebox(0,0)[r]{\strut{}$5.4$}}%
      \colorrgb{0.00,0.38,0.68}%
      \put(792,4310){\makebox(0,0)[r]{\strut{}$5.6$}}%
      \colorrgb{0.00,0.38,0.68}%
      \put(792,4837){\makebox(0,0)[r]{\strut{}$5.8$}}%
      \csname LTb\endcsname%
      \put(924,3035){\makebox(0,0){\strut{}$-100$}}%
      \put(1713,3035){\makebox(0,0){\strut{}$-50$}}%
      \put(2503,3035){\makebox(0,0){\strut{}$0$}}%
      \put(3292,3035){\makebox(0,0){\strut{}$50$}}%
      \put(4081,3035){\makebox(0,0){\strut{}$100$}}%
      \colorrgb{0.87,0.09,0.12}%
      \put(4213,3255){\makebox(0,0)[l]{\strut{}$-3$}}%
      \colorrgb{0.87,0.09,0.12}%
      \put(4213,3782){\makebox(0,0)[l]{\strut{}$-1$}}%
      \colorrgb{0.87,0.09,0.12}%
      \put(4213,4310){\makebox(0,0)[l]{\strut{}$1$}}%
      \colorrgb{0.87,0.09,0.12}%
      \put(4213,4837){\makebox(0,0)[l]{\strut{}$3$}}%
      \csname LTb\endcsname%
      \put(1082,4652){\makebox(0,0)[l]{\strut{}$\text{Gd}_{21}(\text{Fe}_{90}\text{Co}_{10})_{79}(\SI{20}{\nano\meter})$}}%
      \put(2187,3440){\makebox(0,0)[l]{\strut{}$j_\text{HM}=\SI{5.09e6}{\ampere/\centi\meter}$}}%
      \put(49,4846){\makebox(0,0)[l]{\strut{}(a)}}%
    }%
    \gplgaddtomacro\gplfronttext{%
      \colorrgb{0.00,0.38,0.68}%
      \put(154,4046){\rotatebox{-270}{\makebox(0,0){\strut{}$R^{1\omega}_\text{AHE}$ $(\si{\ohm})$}}}%
      \colorrgb{0.87,0.09,0.12}%
      \put(4718,4046){\rotatebox{-270}{\makebox(0,0){\strut{}$R^{2\omega}_\text{AHE}$ $(\si{\milli\ohm})$}}}%
      \csname LTb\endcsname%
      \put(2502,2705){\makebox(0,0){\strut{}$B_x$ $(\si{\milli\tesla})$}}%
    }%
    \gplgaddtomacro\gplbacktext{%
      \csname LTb\endcsname%
      \put(792,704){\makebox(0,0)[r]{\strut{}$-3$}}%
      \put(792,1232){\makebox(0,0)[r]{\strut{}$-2$}}%
      \put(792,1759){\makebox(0,0)[r]{\strut{}$-1$}}%
      \put(792,2287){\makebox(0,0)[r]{\strut{}$0$}}%
      \put(924,484){\makebox(0,0){\strut{}$0$}}%
      \put(1450,484){\makebox(0,0){\strut{}$1$}}%
      \put(1976,484){\makebox(0,0){\strut{}$2$}}%
      \put(2503,484){\makebox(0,0){\strut{}$3$}}%
      \put(3029,484){\makebox(0,0){\strut{}$4$}}%
      \put(3555,484){\makebox(0,0){\strut{}$5$}}%
      \put(4081,484){\makebox(0,0){\strut{}$6$}}%
      \put(2292,2129){\makebox(0,0)[l]{\strut{}$\text{Gd}_{21}(\text{Fe}_{90}\text{Co}_{10})_{79}(20)$}}%
      \put(1503,1284){\makebox(0,0)[l]{\strut{}$t$ in $\si{\nano\meter}$}}%
      \put(49,2295){\makebox(0,0)[l]{\strut{}(b)}}%
    }%
    \gplgaddtomacro\gplfronttext{%
      \csname LTb\endcsname%
      \put(286,1495){\rotatebox{-270}{\makebox(0,0){\strut{}$\mu_0 H'_\text{SL}$ (mT)}}}%
      \colorrgb{0.87,0.09,0.12}%
      \put(4300,1495){\rotatebox{-270}{\makebox(0,0){\strut{}}}}%
      \csname LTb\endcsname%
      \put(2502,154){\makebox(0,0){\strut{}$j_\text{HM}$ ($\SI{e6}{\ampere/\centi\meter}$)}}%
      \csname LTb\endcsname%
      \put(1170,910){\makebox(0,0){\strut{}10}}%
      \put(1511,910){\makebox(0,0){\strut{}15}}%
      \put(1852,910){\makebox(0,0){\strut{}20}}%
      \put(2193,910){\makebox(0,0){\strut{}25}}%
      \put(2534,910){\makebox(0,0){\strut{}30}}%
    }%
    \gplbacktext
    \put(0,0){\includegraphics{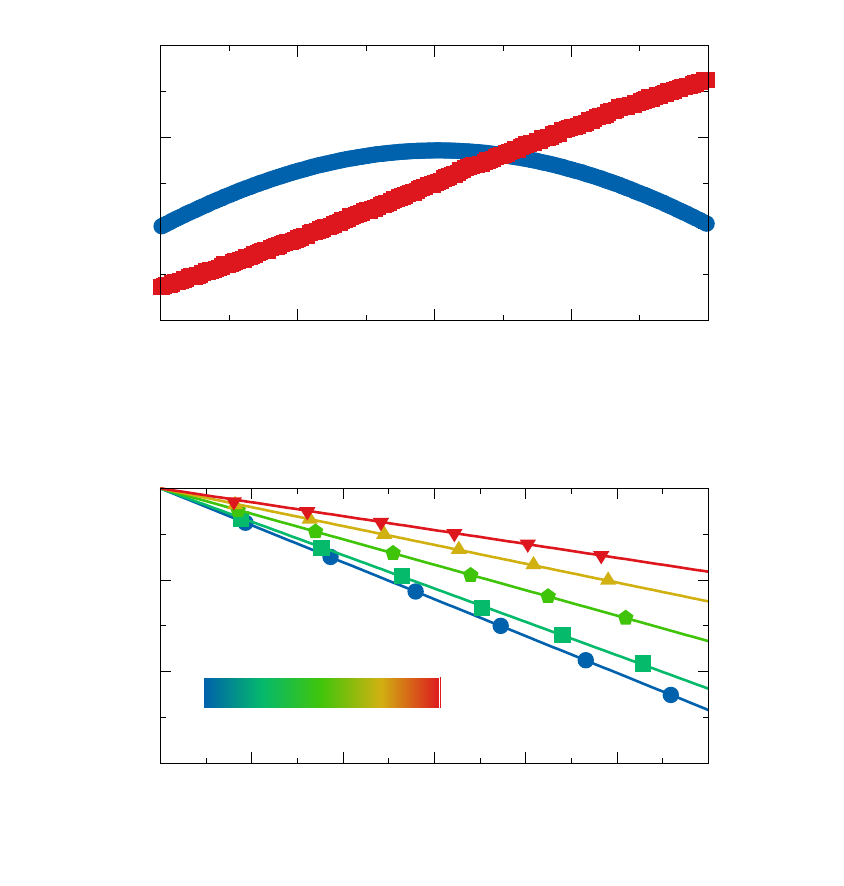}}%
    \gplfronttext
  \end{picture}%
\endgroup
\caption{\label{fig:Harmonic_R2R2wvsH} Harmonic Hall measurement of the effective magnetic field induced by SOT. Panel (a) shows the first- and second harmonic Hall resistance as a function of in-plane magnetic field $B_\text{x}$ parallel to the current direction for a $\SI{20}{\nano\meter}$ thick \GdFeCo film. From this measurement the Slonczewski effective field $\mu_0 H_\text{SL}$ can be extracted. Panel (b) shows that $\mu_0 H'_\text{SL}$ depends linearly on the current density in the heavy metal in all films under investigation. $\mu_0 H'_\text{SL}$ is plotted as measured and not corrected for the planar Hall effect in this figure.}
\end{center}
\end{figure}
The AHE resistivity $\rho_\text{AHE}$ as a function of thickness $t$ is shown in Fig.~\ref{fig:Resistivity}, on the left axis. $\rho_\text{AHE}$ is proportional to the magnetic volume, and thus a linear trend can be observed. In addition, the planar Hall effect (PHE) resistivity $\rho_\text{PHE}$ was measured. $\rho_\text{PHE}$ also follows a linear trend, however, $\rho_\text{PHE}$ is two order of magnitude smaller than $\rho_\text{AHE}$. The ratio of AHE and PHE is $\xi=\rho_\text{PHE}/\rho_\text{AHE}\approx\SI{3.2}{\percent}$. The right axis of Fig.~\ref{fig:Resistivity} shows the saturation magnetization $M_\text s$ as a function of thickness. $M_\text s$ is constant for all samples with $t>\SI{10}{\nano\meter}$. However, the thinnest sample of this series with $t=\SI{10}{\nano\meter}$ has a larger $M_\text s$.

To characterize the SOT in our samples we performed harmonic Hall measurements of the effective magnetic fields, following Hayashi \textit{et al.}~\cite{Hayashi2014}. We start by measuring the first and second harmonic voltage responses $V_{1\omega}$ and $V_{2\omega}$ to an AC current with $\omega=\SI{1.2}{\kilo\hertz}$ as a function of in-plane magnetic field $B_{x,y}$. To measure the Slonczewski field $H_\text{SL}$, the in-plane magnetic field $B_x$ is applied along the $x$-direction, while the in-plane magnetic field $B_y$ is applied along the $y$-direction to measure the field-like field $H_\text{FL}$. For all harmonic Hall measurements discussed in the following, the samples were magnetized along the $+z$-direction and thus $m_\text{tot}>0$.

A typical harmonic Hall measurement to determine $H_\text{SL}$ is shown in Fig.~\ref{fig:Harmonic_R2R2wvsH}(a) for a \GdFeCo sample with $t=\SI{20}{\nano\meter}$ and $m_\text{tot}>0$. The first harmonic response $R^{1\omega}_\text{AHE}=V_{1\omega}/I_\text{FM}$ (blue) follows a quadratic trend as a function of the in-plane magnetic field $B_{x}$, while the second harmonic response $R^{2\omega}_\text{AHE}=V_{2\omega}/I_\text{FM}$ (red) can be approximated by a linear function of $B_x$. Here $I_\text{FM}$ is the current through the \GdFeCo.

\begin{figure}[t]
\begin{center}
%
%
\begingroup
  \makeatletter
  \providecommand\color[2][]{%
    \GenericError{(gnuplot) \space\space\space\@spaces}{%
      Package color not loaded in conjunction with
      terminal option `colourtext'%
    }{See the gnuplot documentation for explanation.%
    }{Either use 'blacktext' in gnuplot or load the package
      color.sty in LaTeX.}%
    \renewcommand\color[2][]{}%
  }%
  \providecommand\includegraphics[2][]{%
    \GenericError{(gnuplot) \space\space\space\@spaces}{%
      Package graphicx or graphics not loaded%
    }{See the gnuplot documentation for explanation.%
    }{The gnuplot epslatex terminal needs graphicx.sty or graphics.sty.}%
    \renewcommand\includegraphics[2][]{}%
  }%
  \providecommand\rotatebox[2]{#2}%
  \@ifundefined{ifGPcolor}{%
    \newif\ifGPcolor
    \GPcolortrue
  }{}%
  \@ifundefined{ifGPblacktext}{%
    \newif\ifGPblacktext
    \GPblacktextfalse
  }{}%
  \let\gplgaddtomacro\g@addto@macro
  \gdef\gplbacktext{}%
  \gdef\gplfronttext{}%
  \makeatother
  \ifGPblacktext
    \def\colorrgb#1{}%
    \def\colorgray#1{}%
  \else
    \ifGPcolor
      \def\colorrgb#1{\color[rgb]{#1}}%
      \def\colorgray#1{\color[gray]{#1}}%
      \expandafter\def\csname LTw\endcsname{\color{white}}%
      \expandafter\def\csname LTb\endcsname{\color{black}}%
      \expandafter\def\csname LTa\endcsname{\color{black}}%
      \expandafter\def\csname LT0\endcsname{\color[rgb]{1,0,0}}%
      \expandafter\def\csname LT1\endcsname{\color[rgb]{0,1,0}}%
      \expandafter\def\csname LT2\endcsname{\color[rgb]{0,0,1}}%
      \expandafter\def\csname LT3\endcsname{\color[rgb]{1,0,1}}%
      \expandafter\def\csname LT4\endcsname{\color[rgb]{0,1,1}}%
      \expandafter\def\csname LT5\endcsname{\color[rgb]{1,1,0}}%
      \expandafter\def\csname LT6\endcsname{\color[rgb]{0,0,0}}%
      \expandafter\def\csname LT7\endcsname{\color[rgb]{1,0.3,0}}%
      \expandafter\def\csname LT8\endcsname{\color[rgb]{0.5,0.5,0.5}}%
    \else
      \def\colorrgb#1{\color{black}}%
      \def\colorgray#1{\color[gray]{#1}}%
      \expandafter\def\csname LTw\endcsname{\color{white}}%
      \expandafter\def\csname LTb\endcsname{\color{black}}%
      \expandafter\def\csname LTa\endcsname{\color{black}}%
      \expandafter\def\csname LT0\endcsname{\color{black}}%
      \expandafter\def\csname LT1\endcsname{\color{black}}%
      \expandafter\def\csname LT2\endcsname{\color{black}}%
      \expandafter\def\csname LT3\endcsname{\color{black}}%
      \expandafter\def\csname LT4\endcsname{\color{black}}%
      \expandafter\def\csname LT5\endcsname{\color{black}}%
      \expandafter\def\csname LT6\endcsname{\color{black}}%
      \expandafter\def\csname LT7\endcsname{\color{black}}%
      \expandafter\def\csname LT8\endcsname{\color{black}}%
    \fi
  \fi
    \setlength{\unitlength}{0.0500bp}%
    \ifx\gptboxheight\undefined%
      \newlength{\gptboxheight}%
      \newlength{\gptboxwidth}%
      \newsavebox{\gptboxtext}%
    \fi%
    \setlength{\fboxrule}{0.5pt}%
    \setlength{\fboxsep}{1pt}%
\begin{picture}(4874.00,4534.00)%
    \gplgaddtomacro\gplbacktext{%
      \csname LTb\endcsname%
      \put(682,2267){\makebox(0,0)[r]{\strut{}$0$}}%
      \put(682,2696){\makebox(0,0)[r]{\strut{}$3$}}%
      \put(682,3125){\makebox(0,0)[r]{\strut{}$6$}}%
      \put(682,3554){\makebox(0,0)[r]{\strut{}$9$}}%
      \put(682,3983){\makebox(0,0)[r]{\strut{}$12$}}%
      \put(814,2047){\makebox(0,0){\strut{}}}%
      \put(1425,2047){\makebox(0,0){\strut{}}}%
      \put(2035,2047){\makebox(0,0){\strut{}}}%
      \put(2646,2047){\makebox(0,0){\strut{}}}%
      \put(3256,2047){\makebox(0,0){\strut{}}}%
      \put(3867,2047){\makebox(0,0){\strut{}}}%
      \put(4477,2047){\makebox(0,0){\strut{}}}%
    }%
    \gplgaddtomacro\gplfronttext{%
      \csname LTb\endcsname%
      \put(176,3125){\rotatebox{-270}{\makebox(0,0){\strut{}$\frac{\mu_0 H_\text{eff}}{j_\text{HM}}$ $(\frac{\text{mT}}{10^7 \text{A}/\text{cm}^2})$}}}%
      \colorrgb{0.00,0.38,0.68}%
      \put(4005,3810){\makebox(0,0)[r]{\strut{}Slonczewski Torque}}%
      \colorrgb{0.87,0.09,0.12}%
      \put(4005,3590){\makebox(0,0)[r]{\strut{}Field-like Torque }}%
    }%
    \gplgaddtomacro\gplbacktext{%
      \csname LTb\endcsname%
      \put(682,550){\makebox(0,0)[r]{\strut{}$0$}}%
      \put(682,1122){\makebox(0,0)[r]{\strut{}$10$}}%
      \put(682,1695){\makebox(0,0)[r]{\strut{}$20$}}%
      \put(682,2267){\makebox(0,0)[r]{\strut{}}}%
      \put(814,330){\makebox(0,0){\strut{}$5$}}%
      \put(1425,330){\makebox(0,0){\strut{}$10$}}%
      \put(2035,330){\makebox(0,0){\strut{}$15$}}%
      \put(2646,330){\makebox(0,0){\strut{}$20$}}%
      \put(3256,330){\makebox(0,0){\strut{}$25$}}%
      \put(3867,330){\makebox(0,0){\strut{}$30$}}%
      \put(4477,330){\makebox(0,0){\strut{}$35$}}%
      \put(2646,1981){\makebox(0,0)[l]{\strut{}$\text{Gd}_{21}(\text{Fe}_{90}\text{Co}_{10})_{79}(t)$}}%
      \colorrgb{0.00,0.38,0.68}%
      \put(875,1723){\makebox(0,0)[l]{\strut{}$18.7\%$}}%
      \colorrgb{0.87,0.09,0.12}%
      \put(875,962){\makebox(0,0)[l]{\strut{}$5.4\%$}}%
    }%
    \gplgaddtomacro\gplfronttext{%
      \csname LTb\endcsname%
      \put(176,1408){\rotatebox{-270}{\makebox(0,0){\strut{}$\theta_\text{eff} (\%)$}}}%
      \put(2645,0){\makebox(0,0){\strut{}$t$ (nm)}}%
    }%
    \gplbacktext
    \put(0,0){\includegraphics{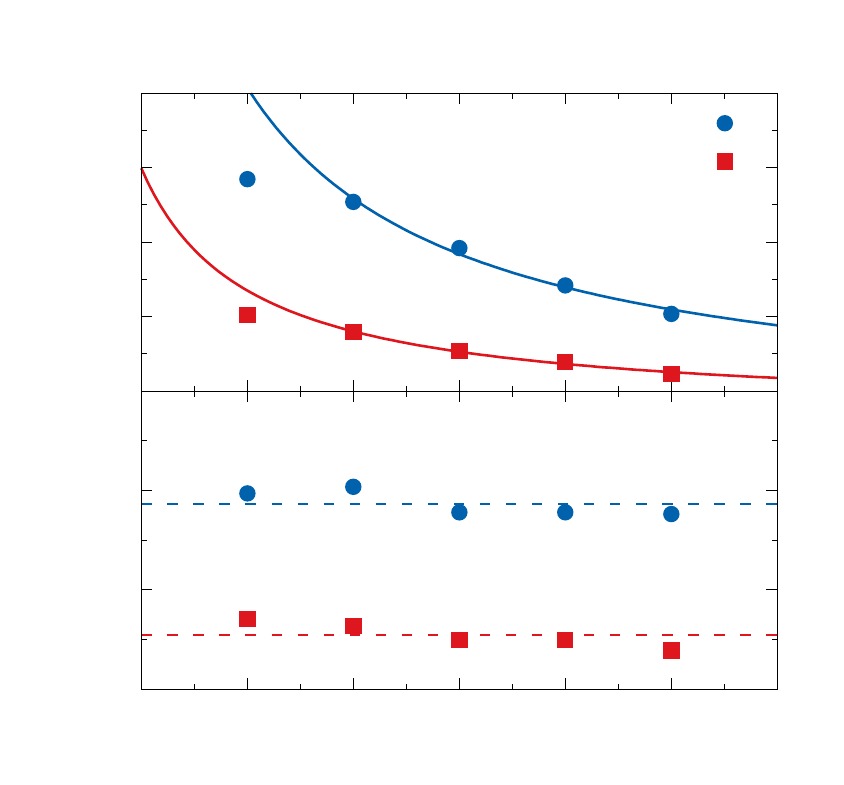}}%
    \gplfronttext
  \end{picture}%
\endgroup
\caption{\label{fig:Heffvst} The upper panel shows the spin torque efficiency as a function of thickness. We find that both the damping-like (blue dots) and the field-like (red squares) torque efficiency decreases linearly with thickness. The lower panel shows the extracted spin Hall angle $\theta_\text{SH}$ for the thickness series. Since the thickness $t$ was varied but the HM layers were kept unchanged (and thus the HM/FM interface is constant), the effective spin Hall angle does not depend on the thickness of the FM.}
\end{center}
\end{figure}
To compute the effective Slonczewski field from this measurement, the slope of $R^{2\omega}_\text{AHE}$ is divided by the curvature of $R^{1\omega}_\text{AHE}$~\cite{Hayashi2014}:
\begin{align}
\mu_0 H'_\text{SL,FL}&=\Bigg(\frac{\partial R^{2\omega}_\text{AHE}}{\partial B_{x,y}}\Bigg)\times\Bigg(\frac{\partial^2 R^{\omega}_\text{AHE}}{\partial B^2_{x,y}}\Bigg)^{-1}\, 
\end{align}
Figure~\ref{fig:Harmonic_R2R2wvsH}(b) shows that the effective magnetic field $\mu_0 H'_\text{SL}$ obtained from this analysis is proportional to the current density in the HM layer in all samples under investigation. This confirms that the effect measured here is indeed caused by SOT, and not by non-linear effects such as heating. In the following, we will refer to the effective field per unit current density through the HM layers as $\zeta_\text{SL,FL}$.

\begin{figure}[t]
\begin{center}
%
%
\begingroup
  \makeatletter
  \providecommand\color[2][]{%
    \GenericError{(gnuplot) \space\space\space\@spaces}{%
      Package color not loaded in conjunction with
      terminal option `colourtext'%
    }{See the gnuplot documentation for explanation.%
    }{Either use 'blacktext' in gnuplot or load the package
      color.sty in LaTeX.}%
    \renewcommand\color[2][]{}%
  }%
  \providecommand\includegraphics[2][]{%
    \GenericError{(gnuplot) \space\space\space\@spaces}{%
      Package graphicx or graphics not loaded%
    }{See the gnuplot documentation for explanation.%
    }{The gnuplot epslatex terminal needs graphicx.sty or graphics.sty.}%
    \renewcommand\includegraphics[2][]{}%
  }%
  \providecommand\rotatebox[2]{#2}%
  \@ifundefined{ifGPcolor}{%
    \newif\ifGPcolor
    \GPcolortrue
  }{}%
  \@ifundefined{ifGPblacktext}{%
    \newif\ifGPblacktext
    \GPblacktextfalse
  }{}%
  \let\gplgaddtomacro\g@addto@macro
  \gdef\gplbacktext{}%
  \gdef\gplfronttext{}%
  \makeatother
  \ifGPblacktext
    \def\colorrgb#1{}%
    \def\colorgray#1{}%
  \else
    \ifGPcolor
      \def\colorrgb#1{\color[rgb]{#1}}%
      \def\colorgray#1{\color[gray]{#1}}%
      \expandafter\def\csname LTw\endcsname{\color{white}}%
      \expandafter\def\csname LTb\endcsname{\color{black}}%
      \expandafter\def\csname LTa\endcsname{\color{black}}%
      \expandafter\def\csname LT0\endcsname{\color[rgb]{1,0,0}}%
      \expandafter\def\csname LT1\endcsname{\color[rgb]{0,1,0}}%
      \expandafter\def\csname LT2\endcsname{\color[rgb]{0,0,1}}%
      \expandafter\def\csname LT3\endcsname{\color[rgb]{1,0,1}}%
      \expandafter\def\csname LT4\endcsname{\color[rgb]{0,1,1}}%
      \expandafter\def\csname LT5\endcsname{\color[rgb]{1,1,0}}%
      \expandafter\def\csname LT6\endcsname{\color[rgb]{0,0,0}}%
      \expandafter\def\csname LT7\endcsname{\color[rgb]{1,0.3,0}}%
      \expandafter\def\csname LT8\endcsname{\color[rgb]{0.5,0.5,0.5}}%
    \else
      \def\colorrgb#1{\color{black}}%
      \def\colorgray#1{\color[gray]{#1}}%
      \expandafter\def\csname LTw\endcsname{\color{white}}%
      \expandafter\def\csname LTb\endcsname{\color{black}}%
      \expandafter\def\csname LTa\endcsname{\color{black}}%
      \expandafter\def\csname LT0\endcsname{\color{black}}%
      \expandafter\def\csname LT1\endcsname{\color{black}}%
      \expandafter\def\csname LT2\endcsname{\color{black}}%
      \expandafter\def\csname LT3\endcsname{\color{black}}%
      \expandafter\def\csname LT4\endcsname{\color{black}}%
      \expandafter\def\csname LT5\endcsname{\color{black}}%
      \expandafter\def\csname LT6\endcsname{\color{black}}%
      \expandafter\def\csname LT7\endcsname{\color{black}}%
      \expandafter\def\csname LT8\endcsname{\color{black}}%
    \fi
  \fi
    \setlength{\unitlength}{0.0500bp}%
    \ifx\gptboxheight\undefined%
      \newlength{\gptboxheight}%
      \newlength{\gptboxwidth}%
      \newsavebox{\gptboxtext}%
    \fi%
    \setlength{\fboxrule}{0.5pt}%
    \setlength{\fboxsep}{1pt}%
\begin{picture}(4874.00,5102.00)%
    \gplgaddtomacro\gplbacktext{%
      \csname LTb\endcsname%
      \put(792,3255){\makebox(0,0)[r]{\strut{}$-2$}}%
      \put(792,4046){\makebox(0,0)[r]{\strut{}$0$}}%
      \put(792,4837){\makebox(0,0)[r]{\strut{}$2$}}%
      \put(924,3035){\makebox(0,0){\strut{}$0$}}%
      \put(1555,3035){\makebox(0,0){\strut{}$20$}}%
      \put(2187,3035){\makebox(0,0){\strut{}$40$}}%
      \put(2818,3035){\makebox(0,0){\strut{}$60$}}%
      \put(3450,3035){\makebox(0,0){\strut{}$80$}}%
      \put(4081,3035){\makebox(0,0){\strut{}$100$}}%
      \put(1997,4639){\makebox(0,0)[l]{\strut{}$\text{Gd}_{21}(\text{Fe}_{90}\text{Co}_{10})_{79}(\SI{10}{\nano\meter})$}}%
      \colorrgb{0.02,0.73,0.42}%
      \put(2187,4125){\makebox(0,0)[l]{\strut{}down $\rightarrow$ up}}%
      \colorrgb{0.82,0.69,0.07}%
      \put(2187,3571){\makebox(0,0)[l]{\strut{}up $\rightarrow$ down}}%
      \csname LTb\endcsname%
      \put(49,4846){\makebox(0,0)[l]{\strut{}(a)}}%
    }%
    \gplgaddtomacro\gplfronttext{%
      \csname LTb\endcsname%
      \put(286,4046){\rotatebox{-270}{\makebox(0,0){\strut{}$j_\text{c}^\text{HM}$ ($10^7$ A/cm$^2$)}}}%
      \put(2502,2705){\makebox(0,0){\strut{}$B_\text{x}$ (mT)}}%
    }%
    \gplgaddtomacro\gplbacktext{%
      \colorrgb{0.00,0.38,0.68}%
      \put(792,704){\makebox(0,0)[r]{\strut{}$0$}}%
      \colorrgb{0.00,0.38,0.68}%
      \put(792,1232){\makebox(0,0)[r]{\strut{}$2$}}%
      \colorrgb{0.00,0.38,0.68}%
      \put(792,1759){\makebox(0,0)[r]{\strut{}$4$}}%
      \colorrgb{0.00,0.38,0.68}%
      \put(792,2287){\makebox(0,0)[r]{\strut{}$6$}}%
      \csname LTb\endcsname%
      \put(924,484){\makebox(0,0){\strut{}$5$}}%
      \put(1450,484){\makebox(0,0){\strut{}$10$}}%
      \put(1976,484){\makebox(0,0){\strut{}$15$}}%
      \put(2503,484){\makebox(0,0){\strut{}$20$}}%
      \put(3029,484){\makebox(0,0){\strut{}$25$}}%
      \put(3555,484){\makebox(0,0){\strut{}$30$}}%
      \put(4081,484){\makebox(0,0){\strut{}$35$}}%
      \colorrgb{0.87,0.09,0.12}%
      \put(4213,704){\makebox(0,0)[l]{\strut{}$0$}}%
      \colorrgb{0.87,0.09,0.12}%
      \put(4213,1232){\makebox(0,0)[l]{\strut{}$2$}}%
      \colorrgb{0.87,0.09,0.12}%
      \put(4213,1759){\makebox(0,0)[l]{\strut{}$4$}}%
      \colorrgb{0.87,0.09,0.12}%
      \put(4213,2287){\makebox(0,0)[l]{\strut{}$6$}}%
      \csname LTb\endcsname%
      \put(2397,889){\makebox(0,0)[l]{\strut{}$\text{Gd}_{21}(\text{Fe}_{90}\text{Co}_{10})_{79}(t)$}}%
      \put(977,2102){\makebox(0,0)[l]{\strut{}$B_\text{IP}=\SI{100}{\milli\tesla}$}}%
      \put(49,2295){\makebox(0,0)[l]{\strut{}(b)}}%
    }%
    \gplgaddtomacro\gplfronttext{%
      \colorrgb{0.00,0.38,0.68}%
      \put(418,1495){\rotatebox{-270}{\makebox(0,0){\strut{}$j_\text{c}^\text{HM}$ ($10^7$ A/cm$^2$)}}}%
      \colorrgb{0.87,0.09,0.12}%
      \put(4586,1495){\rotatebox{-270}{\makebox(0,0){\strut{}$j_\text{c}/\Delta$ ($10^5$ A/cm$^2$)}}}%
      \csname LTb\endcsname%
      \put(2502,154){\makebox(0,0){\strut{}$t$ (nm)}}%
    }%
    \gplbacktext
    \put(0,0){\includegraphics{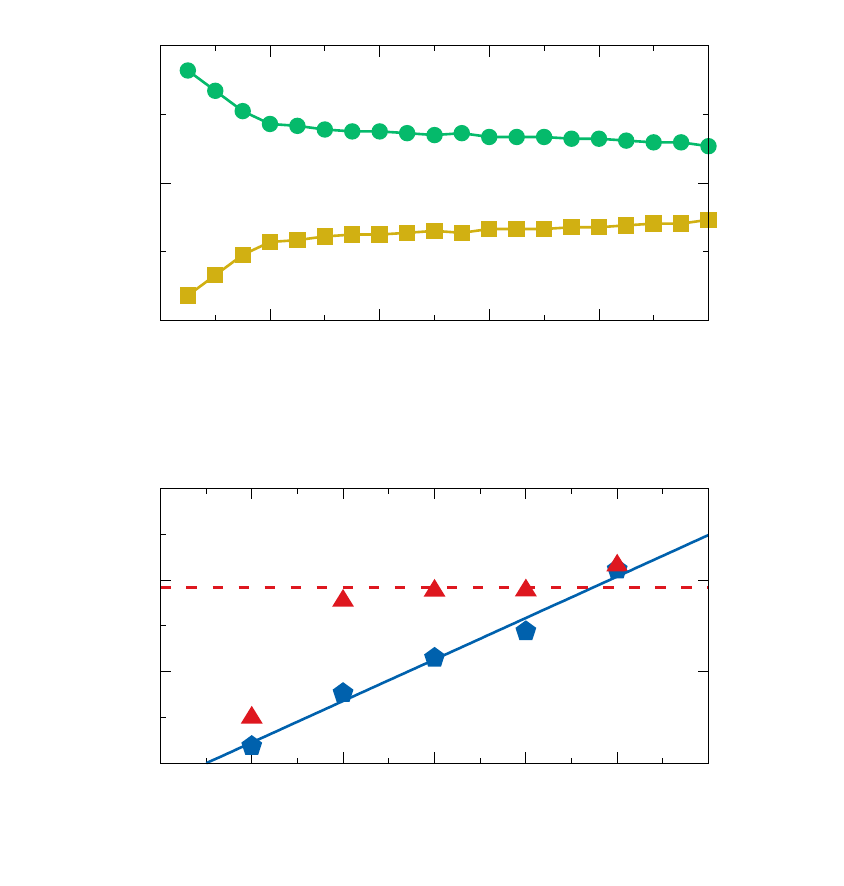}}%
    \gplfronttext
  \end{picture}%
\endgroup
\caption{\label{fig:SPD} Panel (a) shows the switching phase diagram for a \GdFeCo film with thickness $t=\SI{10}{\nano\meter}$. Panel (b) shows that the critical switching current density depends linearly on the film thickness (blue diamonds, left axis). The ratio $j_\text c/\Delta$ is constant for samples with $t>\SI{10}{\nano\meter}$ (red triangles, right axis).}
\end{center}
\end{figure}
\begin{figure*}[t]
\begin{center}
%
%
\begingroup
  \makeatletter
  \providecommand\color[2][]{%
    \GenericError{(gnuplot) \space\space\space\@spaces}{%
      Package color not loaded in conjunction with
      terminal option `colourtext'%
    }{See the gnuplot documentation for explanation.%
    }{Either use 'blacktext' in gnuplot or load the package
      color.sty in LaTeX.}%
    \renewcommand\color[2][]{}%
  }%
  \providecommand\includegraphics[2][]{%
    \GenericError{(gnuplot) \space\space\space\@spaces}{%
      Package graphicx or graphics not loaded%
    }{See the gnuplot documentation for explanation.%
    }{The gnuplot epslatex terminal needs graphicx.sty or graphics.sty.}%
    \renewcommand\includegraphics[2][]{}%
  }%
  \providecommand\rotatebox[2]{#2}%
  \@ifundefined{ifGPcolor}{%
    \newif\ifGPcolor
    \GPcolortrue
  }{}%
  \@ifundefined{ifGPblacktext}{%
    \newif\ifGPblacktext
    \GPblacktextfalse
  }{}%
  \let\gplgaddtomacro\g@addto@macro
  \gdef\gplbacktext{}%
  \gdef\gplfronttext{}%
  \makeatother
  \ifGPblacktext
    \def\colorrgb#1{}%
    \def\colorgray#1{}%
  \else
    \ifGPcolor
      \def\colorrgb#1{\color[rgb]{#1}}%
      \def\colorgray#1{\color[gray]{#1}}%
      \expandafter\def\csname LTw\endcsname{\color{white}}%
      \expandafter\def\csname LTb\endcsname{\color{black}}%
      \expandafter\def\csname LTa\endcsname{\color{black}}%
      \expandafter\def\csname LT0\endcsname{\color[rgb]{1,0,0}}%
      \expandafter\def\csname LT1\endcsname{\color[rgb]{0,1,0}}%
      \expandafter\def\csname LT2\endcsname{\color[rgb]{0,0,1}}%
      \expandafter\def\csname LT3\endcsname{\color[rgb]{1,0,1}}%
      \expandafter\def\csname LT4\endcsname{\color[rgb]{0,1,1}}%
      \expandafter\def\csname LT5\endcsname{\color[rgb]{1,1,0}}%
      \expandafter\def\csname LT6\endcsname{\color[rgb]{0,0,0}}%
      \expandafter\def\csname LT7\endcsname{\color[rgb]{1,0.3,0}}%
      \expandafter\def\csname LT8\endcsname{\color[rgb]{0.5,0.5,0.5}}%
    \else
      \def\colorrgb#1{\color{black}}%
      \def\colorgray#1{\color[gray]{#1}}%
      \expandafter\def\csname LTw\endcsname{\color{white}}%
      \expandafter\def\csname LTb\endcsname{\color{black}}%
      \expandafter\def\csname LTa\endcsname{\color{black}}%
      \expandafter\def\csname LT0\endcsname{\color{black}}%
      \expandafter\def\csname LT1\endcsname{\color{black}}%
      \expandafter\def\csname LT2\endcsname{\color{black}}%
      \expandafter\def\csname LT3\endcsname{\color{black}}%
      \expandafter\def\csname LT4\endcsname{\color{black}}%
      \expandafter\def\csname LT5\endcsname{\color{black}}%
      \expandafter\def\csname LT6\endcsname{\color{black}}%
      \expandafter\def\csname LT7\endcsname{\color{black}}%
      \expandafter\def\csname LT8\endcsname{\color{black}}%
    \fi
  \fi
    \setlength{\unitlength}{0.0500bp}%
    \ifx\gptboxheight\undefined%
      \newlength{\gptboxheight}%
      \newlength{\gptboxwidth}%
      \newsavebox{\gptboxtext}%
    \fi%
    \setlength{\fboxrule}{0.5pt}%
    \setlength{\fboxsep}{1pt}%
\begin{picture}(10204.00,3400.00)%
    \gplgaddtomacro\gplbacktext{%
      \csname LTb\endcsname%
      \put(682,704){\makebox(0,0)[r]{\strut{}$-6$}}%
      \put(682,1312){\makebox(0,0)[r]{\strut{}$-3$}}%
      \put(682,1920){\makebox(0,0)[r]{\strut{}$0$}}%
      \put(682,2527){\makebox(0,0)[r]{\strut{}$3$}}%
      \put(682,3135){\makebox(0,0)[r]{\strut{}$6$}}%
      \put(814,484){\makebox(0,0){\strut{}-15}}%
      \put(1362,484){\makebox(0,0){\strut{}-10}}%
      \put(1910,484){\makebox(0,0){\strut{}}}%
      \put(2457,484){\makebox(0,0){\strut{}10}}%
      \put(3005,484){\makebox(0,0){\strut{}15}}%
      \put(1020,3229){\makebox(0,0)[l]{\strut{}$\text{Gd}_{21}(\text{Fe}_{90}\text{Co}_{10})_{79}(25\text{nm})$}}%
      \colorrgb{0.00,0.38,0.68}%
      \put(2238,1514){\rotatebox{90}{\makebox(0,0)[l]{\strut{}$0.89$ mT/s}}}%
      \colorrgb{0.02,0.73,0.42}%
      \put(2676,1514){\rotatebox{90}{\makebox(0,0)[l]{\strut{}$119$ mT/s}}}%
      \csname LTb\endcsname%
      \put(0,3229){\makebox(0,0)[l]{\strut{}(a)}}%
    }%
    \gplgaddtomacro\gplfronttext{%
      \csname LTb\endcsname%
      \put(176,1919){\rotatebox{-270}{\makebox(0,0){\strut{}$R_\text{AHE}$ $(\si{\ohm})$}}}%
      \put(1909,154){\makebox(0,0){\strut{}$B_\text z$ $(\si{\milli\tesla})$}}%
    }%
    \gplgaddtomacro\gplbacktext{%
      \csname LTb\endcsname%
      \put(4083,704){\makebox(0,0)[r]{\strut{}$0$}}%
      \put(4083,1514){\makebox(0,0)[r]{\strut{}$5$}}%
      \put(4083,2325){\makebox(0,0)[r]{\strut{}$10$}}%
      \put(4083,3135){\makebox(0,0)[r]{\strut{}$15$}}%
      \put(4215,484){\makebox(0,0){\strut{}$0.1$}}%
      \put(4763,484){\makebox(0,0){\strut{}$1$}}%
      \put(5311,484){\makebox(0,0){\strut{}$10$}}%
      \put(5858,484){\makebox(0,0){\strut{}$100$}}%
      \put(6406,484){\makebox(0,0){\strut{}$1000$}}%
      \colorrgb{0.00,0.38,0.68}%
      \put(4763,1158){\rotatebox{10}{\makebox(0,0)[l]{\strut{}$t=10$nm}}}%
      \colorrgb{0.02,0.73,0.42}%
      \put(4763,1482){\rotatebox{10}{\makebox(0,0)[l]{\strut{}$t=15$nm}}}%
      \colorrgb{0.25,0.77,0.04}%
      \put(4763,1920){\rotatebox{10}{\makebox(0,0)[l]{\strut{}$t=20$nm}}}%
      \colorrgb{0.82,0.69,0.07}%
      \put(4763,2325){\rotatebox{10}{\makebox(0,0)[l]{\strut{}$t=25$nm}}}%
      \colorrgb{0.87,0.09,0.12}%
      \put(4763,2568){\rotatebox{10}{\makebox(0,0)[l]{\strut{}$t=30$nm}}}%
      \csname LTb\endcsname%
      \put(4795,3229){\makebox(0,0)[l]{\strut{}$\text{Gd}_{21}(\text{Fe}_{90}\text{Co}_{10})_{79}(t)$}}%
      \put(3469,3229){\makebox(0,0)[l]{\strut{}(b)}}%
    }%
    \gplgaddtomacro\gplfronttext{%
      \csname LTb\endcsname%
      \put(3577,1919){\rotatebox{-270}{\makebox(0,0){\strut{}$\mu_0H_\text c$ $(\si{\milli\tesla})$}}}%
      \put(5310,154){\makebox(0,0){\strut{}rate $(\si{\milli\tesla}/\si{\second})$}}%
    }%
    \gplgaddtomacro\gplbacktext{%
      \csname LTb\endcsname%
      \put(7484,704){\makebox(0,0)[r]{\strut{}$25$}}%
      \put(7484,1514){\makebox(0,0)[r]{\strut{}$50$}}%
      \put(7484,2325){\makebox(0,0)[r]{\strut{}$75$}}%
      \put(7484,3135){\makebox(0,0)[r]{\strut{}$100$}}%
      \put(7616,484){\makebox(0,0){\strut{}$5$}}%
      \put(7981,484){\makebox(0,0){\strut{}$10$}}%
      \put(8346,484){\makebox(0,0){\strut{}$15$}}%
      \put(8712,484){\makebox(0,0){\strut{}$20$}}%
      \put(9077,484){\makebox(0,0){\strut{}$25$}}%
      \put(9442,484){\makebox(0,0){\strut{}$30$}}%
      \put(9807,484){\makebox(0,0){\strut{}$35$}}%
      \put(8162,3229){\makebox(0,0)[l]{\strut{}$\text{Gd}_{21}(\text{Fe}_{90}\text{Co}_{10})_{79}(t)$}}%
      \put(6836,3229){\makebox(0,0)[l]{\strut{}(c)}}%
    }%
    \gplgaddtomacro\gplfronttext{%
      \csname LTb\endcsname%
      \put(6978,1919){\rotatebox{-270}{\makebox(0,0){\strut{}$\Delta$}}}%
      \put(8711,154){\makebox(0,0){\strut{}$t$ $(\si{\nano\meter})$}}%
    }%
    \gplbacktext
    \put(0,0){\includegraphics{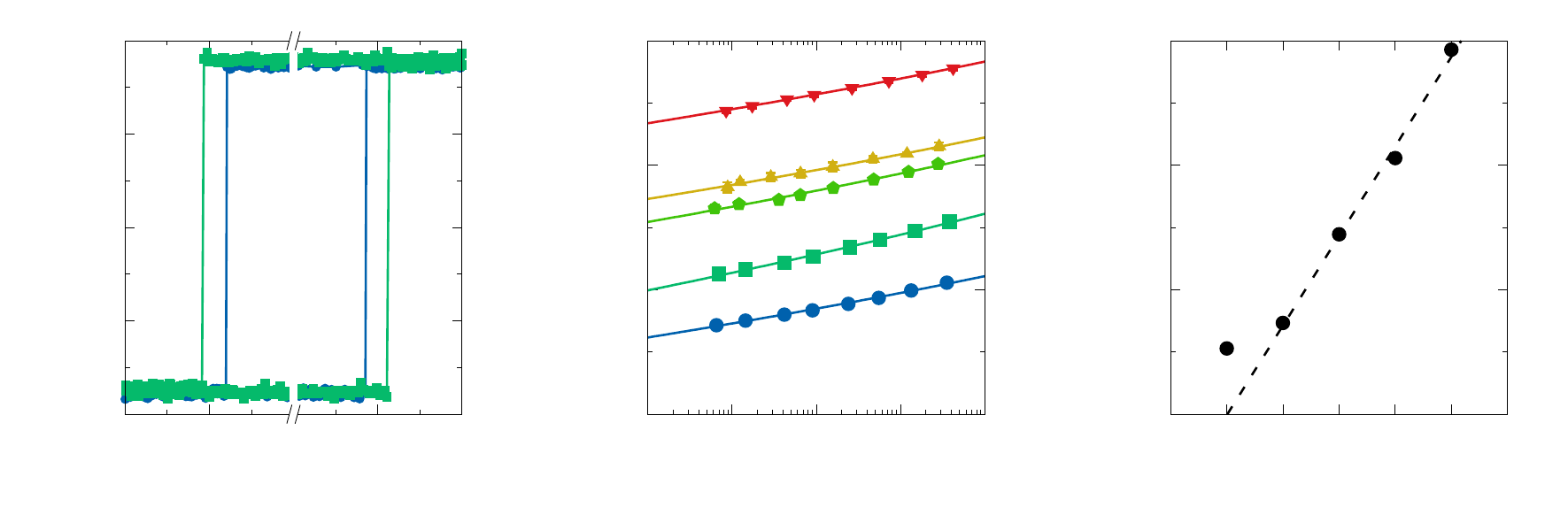}}%
    \gplfronttext
  \end{picture}%
\endgroup
\caption{\label{fig:ThermalStability} Measurement of the thermal stability $\Delta$. Panel (a) shows typical hysteresis loops for the films under investigation. $B_\text z$ is applied along the easy axis of the magnet. The coercive field $\mu_0H_\text C$ is increasing with the sweep rate of the magnetic field due to the thermal nature of the switching events. Panel (b) shows the dependence of $\mu_0H_\text C$ on the magnetic field sweep rate. The solid lines are fits to the Chantrell model, which is used to determine the thermal stability factor $\Delta$. The extracted values for $\Delta$ are plotted in panel (c) as a function of \GdFeCo thickness.}
\end{center}
\end{figure*}
To account for a mixing of $H'_\text{SL}$ and $H'_\text{FL}$ due to the presence of the PHE, the following correction is applied to the data shown in Figure~\ref{fig:Harmonic_R2R2wvsH}~\cite{Hayashi2014}:
\begin{align}
H_\text{SL,FL}&=-2 \frac{H'_\text{SL,FL}+ 2 \xi H'_\text{FL,SL}}{1-4\xi^2}\label{eq:EffectiveField}\, .
\end{align}
Here, $\xi$ is the ratio of PHE resistivity and AHE resistivity: $\xi=\rho_\text{PHE}/\rho_\text{AHE}$ and the sample is magnetized upwards $(m_\text{tot}>0)$. The resulting fields $\zeta_\text{SL,FL}$ are plotted in the upper panel of Fig.~\ref{fig:Heffvst} as a function of \GdFeCo thickness. All samples with $t>\SI{10}{\nano\meter}$, follow a $1/t$ dependence. This is expected from angular momentum conservation as the SOT due to a spin-current is of the form:
\begin{align}
\bm\tau_\text{SOT}=\frac{\hbar}{2e}\frac{j_\text s}{M_\text s t}\bm m\times(\bm\sigma\times \bm m)\, ,\label{eq:torque}
\end{align}
where $e$ is the electric charge, $\hbar$ Planck's constant and $j_s$ the spin current. The sample with $t=\SI{10}{\nano\meter}$ does not follow the trend because $M_\text s$ is larger (\textit{c. f.} Fig.~\ref{fig:Resistivity}) and thus $H_\text{SL,FL}$ is smaller.

Next we use the results for $\zeta_\text{eff}$ to calculate the spin-Hall angle $\theta_\text{SH}$ according to:
\begin{align}
\theta_\text{SH}=\frac{2|e|}{\hbar}M_\text s t_\text{FM} \frac{\mu_0 H_\text{SL,FL}}{j_\text{HM}}\, .\label{eq:SHA}
\end{align}
Here perfect interface transparency was assumed. Figure~\ref{fig:Heffvst} shows that $\theta_\text{SH}$ does not depend on $t$. This is expected, because the SHA only depends on the choice of SH metal and the interface. Both did not change upon increasing the \GdFeCo thickness. The average value for the SHA is $\theta_\text{SH}=\SI{18.7}{\percent}$. This value is well in agreement with literature values in Pt/Co/Ta systems~\cite{Woo2014}.

In addition to harmonic Hall measurements, we also conducted SOT switching experiments to measure the scaling of critical switching current. To this end, current pulses were applied through the heavy metal layers and the magnetic response was measured by the AHE in the \GdFeCo. In order to switch a magnetic film with PMA, an additional magnetic field $B_\text{x}$ need to be applied in the $x$-direction to break the symmetry. First we investigated, how the critical current that is needed to switch the magnet, scales with $B_{x}$. A typical switching phase diagram for a \GdFeCo film with $t=\SI{10}{\nano\meter}$ is shown in Fig.~\ref{fig:SPD}(a). As expected, the critical current decreases with increasing $B_\text{x}$.

Next we show that the critical current density for SOT-switching $j_\text c$ at fixed $B_\text{x}=\SI{100}{\milli\tesla}$ scales linearly with $t$ (blue pentagons in Fig.~\ref{fig:SPD}(b)). This is expected from macro spin simulations if $B_\text{x}< \mu_0 H_\text k$~\cite{Lee2013}, where $H_\text k$ is the anisotropy field. The magnitude of $j_\text c$ is in agreement with previous results in GdFeCo~\cite{Roschewsky2016}.

Magnetic memory applications require not only low switching current densities but also high thermal stability $\Delta=E_\text B/k_\text B T$, where $E_B$ is the activation energy barrier. To investigate the scaling behavior of $\Delta$ with $t$ we performed field-switching experiments with an external magnetic field $B_\text z$ applied along the magnetic easy axis. Field switching is a thermally activated process. Thus, it is expected that the coercive field $H_\text c$ depends on the rate at which the magnetic field is changed. This is shown in Fig.~\ref{fig:ThermalStability}(a), where two hysteresis loops, recorded at different sweep rates, are shown. At slow sweeping rates, the switching event occurs at smaller fields, as it is more likely to thermally nucleate a domain, which will then propagate in the magnet.

The dependence of $H_\text c$ on the sweeping rate has been studied systematically over 3 orders of magnitude on samples with different GdFeCo thickness. The result is shown in Fig.~\ref{fig:ThermalStability}(b): In a semilogarithmic plot $H_\text c$ depends approximately linearly on the sweeping rate. Note that each value for $H_\text c$ is the average value from 20 consecutive field switching measurements. To analyze this data we use a model proposed by El-Hilo \textit{et al.}~\cite{ElHilo1992}:
\begin{align}
H_\text c=H_\text k\left(1-\sqrt{\frac{1}{\Delta}\ln\left[\frac{f_0 H_\text k}{2\Delta}\frac{1}{r}\right]}\right)\, .
\end{align}
Here, $H_\text k$ is the anisotropy field, $\Delta$ is the thermal stability factor, $f_0=\SI{1e11}{\giga\hertz}$ the attempt frequency and $r$ the sweeping rate. Figure~\ref{fig:ThermalStability}(b) shows that this model (solid lines) fits the experimental data well. The extracted values for $\Delta$ are shown in Fig.~\ref{fig:ThermalStability}(c). For the extraction we assumed an attempt frequency of $f_0=\SI{1e10}{\hertz}$.

A linear trend is seen in Fig.~\ref{fig:ThermalStability}(c) because the thermal stability is proportional to the magnetic volume. The sample with $t=\SI{10}{\nano\meter}$ does not follow the trend, which we attribute again to an increase in saturation magnetization.

The switching efficiency for SOT devices is defined as $j_\text c/\Delta$. In our devices, $j_\text c$ as well as $\Delta$ scale linearly with thickness. Thus, the switching efficiency does not depend on the thickness as shown in Fig.~\ref{fig:SPD}(b) on the right axis (red triangles). This has important consequences for memory applications. If traditional SOT devices with TM magnets are scaled to smaller lateral dimensions, the thermal stability will necessarily decrease. Thus, materials with larger anisotropy have to be found to keep the data retention time large. In our devices, on the other hand, the thermal stability can simply be enhanced by increasing the thickness. Thus GdFeCo might be a good material for ultra-scaled memory devices.

All results discussed to far have been obtained on Gd$_x($Fe$_{90}$Co$_{10})_{100-x}$ samples with $x=\SI{21}{\percent}$. However, it is known from previous experiments in RE-TM magnets~\cite{Ueda2016,Roschewsky2016,Finley2016}, that SOT exhibit a distinctive dependence on the film composition. So far the composition dependence has only been studied in very thin films with $t\le\SI{5}{\nano\meter}$. Further, the effective fields were estimated from domain wall motion experiments. Here we used harmonic Hall measurements to characterize the composition dependence of the effective SOT fields in $\SI{30}{\nano\meter}$ thick Gd$_x($Fe$_{90}$Co$_{10})_{100-x}$ films. 

\begin{figure}
\begin{center}
%
%
\begingroup
  \makeatletter
  \providecommand\color[2][]{%
    \GenericError{(gnuplot) \space\space\space\@spaces}{%
      Package color not loaded in conjunction with
      terminal option `colourtext'%
    }{See the gnuplot documentation for explanation.%
    }{Either use 'blacktext' in gnuplot or load the package
      color.sty in LaTeX.}%
    \renewcommand\color[2][]{}%
  }%
  \providecommand\includegraphics[2][]{%
    \GenericError{(gnuplot) \space\space\space\@spaces}{%
      Package graphicx or graphics not loaded%
    }{See the gnuplot documentation for explanation.%
    }{The gnuplot epslatex terminal needs graphicx.sty or graphics.sty.}%
    \renewcommand\includegraphics[2][]{}%
  }%
  \providecommand\rotatebox[2]{#2}%
  \@ifundefined{ifGPcolor}{%
    \newif\ifGPcolor
    \GPcolortrue
  }{}%
  \@ifundefined{ifGPblacktext}{%
    \newif\ifGPblacktext
    \GPblacktextfalse
  }{}%
  \let\gplgaddtomacro\g@addto@macro
  \gdef\gplbacktext{}%
  \gdef\gplfronttext{}%
  \makeatother
  \ifGPblacktext
    \def\colorrgb#1{}%
    \def\colorgray#1{}%
  \else
    \ifGPcolor
      \def\colorrgb#1{\color[rgb]{#1}}%
      \def\colorgray#1{\color[gray]{#1}}%
      \expandafter\def\csname LTw\endcsname{\color{white}}%
      \expandafter\def\csname LTb\endcsname{\color{black}}%
      \expandafter\def\csname LTa\endcsname{\color{black}}%
      \expandafter\def\csname LT0\endcsname{\color[rgb]{1,0,0}}%
      \expandafter\def\csname LT1\endcsname{\color[rgb]{0,1,0}}%
      \expandafter\def\csname LT2\endcsname{\color[rgb]{0,0,1}}%
      \expandafter\def\csname LT3\endcsname{\color[rgb]{1,0,1}}%
      \expandafter\def\csname LT4\endcsname{\color[rgb]{0,1,1}}%
      \expandafter\def\csname LT5\endcsname{\color[rgb]{1,1,0}}%
      \expandafter\def\csname LT6\endcsname{\color[rgb]{0,0,0}}%
      \expandafter\def\csname LT7\endcsname{\color[rgb]{1,0.3,0}}%
      \expandafter\def\csname LT8\endcsname{\color[rgb]{0.5,0.5,0.5}}%
    \else
      \def\colorrgb#1{\color{black}}%
      \def\colorgray#1{\color[gray]{#1}}%
      \expandafter\def\csname LTw\endcsname{\color{white}}%
      \expandafter\def\csname LTb\endcsname{\color{black}}%
      \expandafter\def\csname LTa\endcsname{\color{black}}%
      \expandafter\def\csname LT0\endcsname{\color{black}}%
      \expandafter\def\csname LT1\endcsname{\color{black}}%
      \expandafter\def\csname LT2\endcsname{\color{black}}%
      \expandafter\def\csname LT3\endcsname{\color{black}}%
      \expandafter\def\csname LT4\endcsname{\color{black}}%
      \expandafter\def\csname LT5\endcsname{\color{black}}%
      \expandafter\def\csname LT6\endcsname{\color{black}}%
      \expandafter\def\csname LT7\endcsname{\color{black}}%
      \expandafter\def\csname LT8\endcsname{\color{black}}%
    \fi
  \fi
    \setlength{\unitlength}{0.0500bp}%
    \ifx\gptboxheight\undefined%
      \newlength{\gptboxheight}%
      \newlength{\gptboxwidth}%
      \newsavebox{\gptboxtext}%
    \fi%
    \setlength{\fboxrule}{0.5pt}%
    \setlength{\fboxsep}{1pt}%
\begin{picture}(4874.00,4534.00)%
    \gplgaddtomacro\gplbacktext{%
      \csname LTb\endcsname%
      \put(682,2267){\makebox(0,0)[r]{\strut{}$0$}}%
      \put(682,2839){\makebox(0,0)[r]{\strut{}$4$}}%
      \put(682,3411){\makebox(0,0)[r]{\strut{}$8$}}%
      \put(682,3983){\makebox(0,0)[r]{\strut{}$12$}}%
      \put(814,2047){\makebox(0,0){\strut{}}}%
      \put(1337,2047){\makebox(0,0){\strut{}}}%
      \put(1861,2047){\makebox(0,0){\strut{}}}%
      \put(2384,2047){\makebox(0,0){\strut{}}}%
      \put(2907,2047){\makebox(0,0){\strut{}}}%
      \put(3430,2047){\makebox(0,0){\strut{}}}%
      \put(3954,2047){\makebox(0,0){\strut{}}}%
      \put(4477,2047){\makebox(0,0){\strut{}}}%
      \put(2227,4126){\makebox(0,0)[l]{\strut{}$\text{Gd}_{x}(\text{Fe}_{90}\text{Co}_{10})_{100-x}(30\text{nm})$}}%
      \colorrgb{0.02,0.73,0.42}%
      \put(2750,3697){\makebox(0,0)[l]{\strut{}compensation point}}%
    }%
    \gplgaddtomacro\gplfronttext{%
      \csname LTb\endcsname%
      \put(176,3125){\rotatebox{-270}{\makebox(0,0){\strut{}$\frac{\mu_0 H_\text{eff}}{j_\text{HM}}$ $(\frac{\text{mT}}{10^7 \text{A}/\text{cm}^2})$}}}%
    }%
    \gplgaddtomacro\gplbacktext{%
      \csname LTb\endcsname%
      \put(682,550){\makebox(0,0)[r]{\strut{}$0$}}%
      \put(682,1041){\makebox(0,0)[r]{\strut{}$10$}}%
      \put(682,1531){\makebox(0,0)[r]{\strut{}$20$}}%
      \put(682,2022){\makebox(0,0)[r]{\strut{}$30$}}%
      \put(814,330){\makebox(0,0){\strut{}$20$}}%
      \put(1337,330){\makebox(0,0){\strut{}$21$}}%
      \put(1861,330){\makebox(0,0){\strut{}$22$}}%
      \put(2384,330){\makebox(0,0){\strut{}$23$}}%
      \put(2907,330){\makebox(0,0){\strut{}$24$}}%
      \put(3430,330){\makebox(0,0){\strut{}$25$}}%
      \put(3954,330){\makebox(0,0){\strut{}$26$}}%
      \put(4477,330){\makebox(0,0){\strut{}$27$}}%
      \colorrgb{0.00,0.38,0.68}%
      \put(919,1556){\makebox(0,0)[l]{\strut{}$18.5\%$}}%
      \colorrgb{0.87,0.09,0.12}%
      \put(919,854){\makebox(0,0)[l]{\strut{}$3.6\%$}}%
    }%
    \gplgaddtomacro\gplfronttext{%
      \csname LTb\endcsname%
      \put(176,1408){\rotatebox{-270}{\makebox(0,0){\strut{}$\theta_\text{eff} (\%)$}}}%
      \put(2645,0){\makebox(0,0){\strut{}$x$ ($\%$)}}%
      \colorrgb{0.00,0.38,0.68}%
      \put(4137,2059){\makebox(0,0)[r]{\strut{}Slonczewski Torque}}%
      \colorrgb{0.87,0.09,0.12}%
      \put(4137,1839){\makebox(0,0)[r]{\strut{}Field-like Torque}}%
    }%
    \gplbacktext
    \put(0,0){\includegraphics{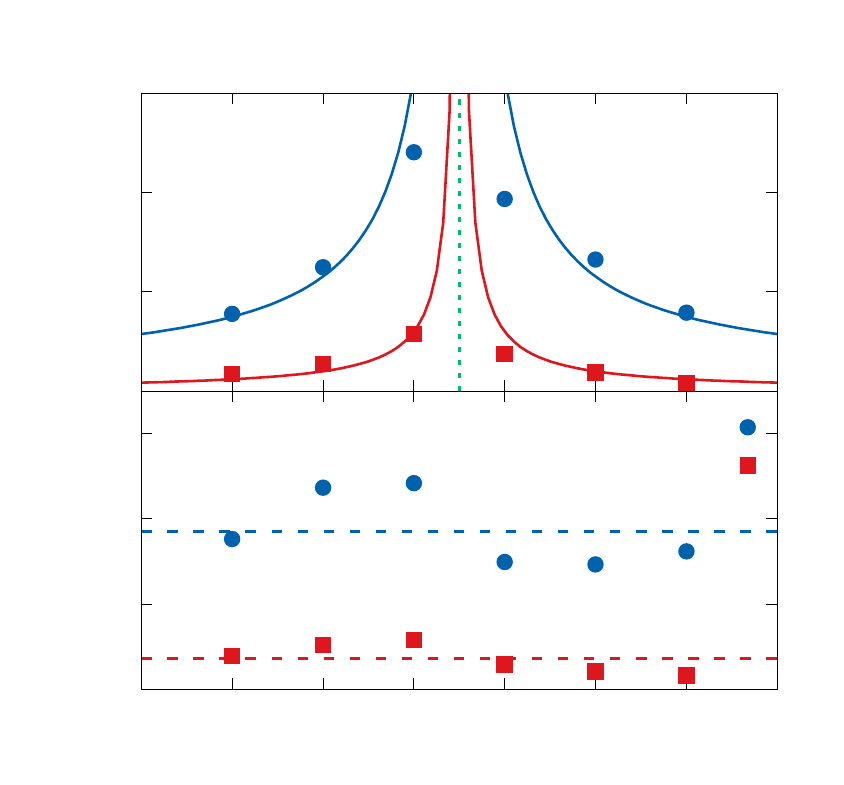}}%
    \gplfronttext
  \end{picture}%
\endgroup
\caption{\label{fig:Heffvsc} Upper panel: The spin torque efficiency as a function of film composition $x$ is shown for the $t=\SI{30}{\nano\meter}$ sample. The spin torque efficiency diverges close to the compensation point. Lower panel: The effective spin Hall angle $\theta_\text{SH}$ is plotted as a function of composition. $\theta_\text{SH}$ shows no distinct dependence on the composition $x$, since the HM/FM interface is unaltered.}
\end{center}
\end{figure}

Following eqn.~\eqref{eq:torque}, we expect the effective magnetic field to diverge at the magnetization compensation point, since $M_\text s$ will vanish. This trend can clearly be seen in Fig.~\ref{fig:Heffvsc}, where we plot $\mu_0 H_\text{SL,FL}/j_\text{HM}$ vs. the concentration $x$ for the damping-like field as well as for the field-like field.

Next, we calculate $\theta_\text{eff}$, using the values obtained for $\mu_0 H_\text{eff}/j_\text{HM}$. Figure~\ref{fig:Heffvsc} shows that $\theta_\text{eff}$ is approximately constant, since only the composition $x$ but not the SH metal or the interfaces were changed. The average value for the SHA is $\theta_\text{eff}=\SI{18.5}{\percent}$. This value is well in agreement with the value of $\theta_\text{eff}$ reported in Fig.~\ref{fig:Heffvst}.

\section{Conclusion}
In conclusion, we showed that the effective magnetic field $H_\text{SL,FL}$, induced by SOT, scales as $H_\text{SL,FL}\propto M^{-1}_\text s t^{-1}_\text{FM}$ for all samples under investigation in this study. This is a consequence of angular momentum conservation. $\theta_\text{SH}$ is found to be constant at an average value of $\theta_\text{SL}=\SI{18}{\percent}$. In addition, we showed SOT-driven switching, even for $\SI{30}{\nano\meter}$ thick magnets. Since $j_\text c$, as well as $\Delta$ scales linearly with thickness, the switching efficiency $j_\text c/\Delta$ is constant. This has important technological implications, as it allows for lateral scaling of memory devices to ultra-small dimensions.

We thank Johannes Mendil and Samuel Smith for fruitful discussions. Research was supported by the Office of Science, Office of Basic Energy Sciences, Materials Science and Engineering Division and the U.S. Department of Energy under Contract No. DE-AC02-05-CH11231 within the NEMM program (KC2204).  Device fabrication was supported by the STARNET/FAME Center.

%
%


\FloatBarrier
\newpage

\bibliography{literature}

\begin{thebibliography}{25}%
\makeatletter
\providecommand \@ifxundefined [1]{%
 \@ifx{#1\undefined}
}%
\providecommand \@ifnum [1]{%
 \ifnum #1\expandafter \@firstoftwo
 \else \expandafter \@secondoftwo
 \fi
}%
\providecommand \@ifx [1]{%
 \ifx #1\expandafter \@firstoftwo
 \else \expandafter \@secondoftwo
 \fi
}%
\providecommand \natexlab [1]{#1}%
\providecommand \enquote  [1]{``#1''}%
\providecommand \bibnamefont  [1]{#1}%
\providecommand \bibfnamefont [1]{#1}%
\providecommand \citenamefont [1]{#1}%
\providecommand \href@noop [0]{\@secondoftwo}%
\providecommand \href [0]{\begingroup \@sanitize@url \@href}%
\providecommand \@href[1]{\@@startlink{#1}\@@href}%
\providecommand \@@href[1]{\endgroup#1\@@endlink}%
\providecommand \@sanitize@url [0]{\catcode `\\12\catcode `\$12\catcode
  `\&12\catcode `\#12\catcode `\^12\catcode `\_12\catcode `\%12\relax}%
\providecommand \@@startlink[1]{}%
\providecommand \@@endlink[0]{}%
\providecommand \url  [0]{\begingroup\@sanitize@url \@url }%
\providecommand \@url [1]{\endgroup\@href {#1}{\urlprefix }}%
\providecommand \urlprefix  [0]{URL }%
\providecommand \Eprint [0]{\href }%
\providecommand \doibase [0]{http://dx.doi.org/}%
\providecommand \selectlanguage [0]{\@gobble}%
\providecommand \bibinfo  [0]{\@secondoftwo}%
\providecommand \bibfield  [0]{\@secondoftwo}%
\providecommand \translation [1]{[#1]}%
\providecommand \BibitemOpen [0]{}%
\providecommand \bibitemStop [0]{}%
\providecommand \bibitemNoStop [0]{.\EOS\space}%
\providecommand \EOS [0]{\spacefactor3000\relax}%
\providecommand \BibitemShut  [1]{\csname bibitem#1\endcsname}%
\let\auto@bib@innerbib\@empty
\bibitem [{\citenamefont {Apalkov}\ \emph {et~al.}(2016)\citenamefont
  {Apalkov}, \citenamefont {Dieny},\ and\ \citenamefont
  {Slaughter}}]{Apalkov2016}%
  \BibitemOpen
  \bibfield  {author} {\bibinfo {author} {\bibfnamefont {D.}~\bibnamefont
  {Apalkov}}, \bibinfo {author} {\bibfnamefont {B.}~\bibnamefont {Dieny}}, \
  and\ \bibinfo {author} {\bibfnamefont {J.~M.}\ \bibnamefont {Slaughter}},\
  }\href {\doibase 10.1109/JPROC.2016.2590142} {\bibfield  {journal} {\bibinfo
  {journal} {Proc. IEEE}\ }\textbf {\bibinfo {volume} {104}},\ \bibinfo {pages}
  {1796} (\bibinfo {year} {2016})}\BibitemShut {NoStop}%
\bibitem [{\citenamefont {Engel}\ \emph {et~al.}(2005)\citenamefont {Engel},
  \citenamefont {Akerman}, \citenamefont {Butcher}, \citenamefont {Dave},
  \citenamefont {DeHerrera}, \citenamefont {Durlam}, \citenamefont
  {Grynkewich}, \citenamefont {Janesky}, \citenamefont {Pietambaram},
  \citenamefont {Rizzo}, \citenamefont {Slaughter}, \citenamefont {Smith},
  \citenamefont {Sun},\ and\ \citenamefont {Tehrani}}]{Engel2005}%
  \BibitemOpen
  \bibfield  {author} {\bibinfo {author} {\bibfnamefont {B.~N.}\ \bibnamefont
  {Engel}}, \bibinfo {author} {\bibfnamefont {J.}~\bibnamefont {Akerman}},
  \bibinfo {author} {\bibfnamefont {B.}~\bibnamefont {Butcher}}, \bibinfo
  {author} {\bibfnamefont {R.~W.}\ \bibnamefont {Dave}}, \bibinfo {author}
  {\bibfnamefont {M.}~\bibnamefont {DeHerrera}}, \bibinfo {author}
  {\bibfnamefont {M.}~\bibnamefont {Durlam}}, \bibinfo {author} {\bibfnamefont
  {G.}~\bibnamefont {Grynkewich}}, \bibinfo {author} {\bibfnamefont
  {J.}~\bibnamefont {Janesky}}, \bibinfo {author} {\bibfnamefont {S.~V.}\
  \bibnamefont {Pietambaram}}, \bibinfo {author} {\bibfnamefont {N.~D.}\
  \bibnamefont {Rizzo}}, \bibinfo {author} {\bibfnamefont {J.~M.}\ \bibnamefont
  {Slaughter}}, \bibinfo {author} {\bibfnamefont {K.}~\bibnamefont {Smith}},
  \bibinfo {author} {\bibfnamefont {J.~J.}\ \bibnamefont {Sun}}, \ and\
  \bibinfo {author} {\bibfnamefont {S.}~\bibnamefont {Tehrani}},\ }\href
  {\doibase 10.1109/TMAG.2004.840847} {\bibfield  {journal} {\bibinfo
  {journal} {IEEE Trans. Mag.}\ }\textbf {\bibinfo {volume} {41}},\ \bibinfo
  {pages} {132} (\bibinfo {year} {2005})}\BibitemShut {NoStop}%
\bibitem [{\citenamefont {Rizzo}\ \emph {et~al.}(2013)\citenamefont {Rizzo},
  \citenamefont {Houssameddine}, \citenamefont {Janesky}, \citenamefont {Whig},
  \citenamefont {Mancoff}, \citenamefont {Schneider}, \citenamefont
  {DeHerrera}, \citenamefont {Sun}, \citenamefont {Nagel}, \citenamefont
  {Deshpande}, \citenamefont {Chia}, \citenamefont {Alam}, \citenamefont
  {Andre}, \citenamefont {Aggarwal},\ and\ \citenamefont
  {Slaughter}}]{Rizzo2013}%
  \BibitemOpen
  \bibfield  {author} {\bibinfo {author} {\bibfnamefont {N.~D.}\ \bibnamefont
  {Rizzo}}, \bibinfo {author} {\bibfnamefont {D.}~\bibnamefont
  {Houssameddine}}, \bibinfo {author} {\bibfnamefont {J.}~\bibnamefont
  {Janesky}}, \bibinfo {author} {\bibfnamefont {R.}~\bibnamefont {Whig}},
  \bibinfo {author} {\bibfnamefont {F.~B.}\ \bibnamefont {Mancoff}}, \bibinfo
  {author} {\bibfnamefont {M.~L.}\ \bibnamefont {Schneider}}, \bibinfo {author}
  {\bibfnamefont {M.}~\bibnamefont {DeHerrera}}, \bibinfo {author}
  {\bibfnamefont {J.~J.}\ \bibnamefont {Sun}}, \bibinfo {author} {\bibfnamefont
  {K.}~\bibnamefont {Nagel}}, \bibinfo {author} {\bibfnamefont
  {S.}~\bibnamefont {Deshpande}}, \bibinfo {author} {\bibfnamefont {H.~J.}\
  \bibnamefont {Chia}}, \bibinfo {author} {\bibfnamefont {S.~M.}\ \bibnamefont
  {Alam}}, \bibinfo {author} {\bibfnamefont {T.}~\bibnamefont {Andre}},
  \bibinfo {author} {\bibfnamefont {S.}~\bibnamefont {Aggarwal}}, \ and\
  \bibinfo {author} {\bibfnamefont {J.~M.}\ \bibnamefont {Slaughter}},\ }\href
  {\doibase 10.1109/TMAG.2013.2243133} {\bibfield  {journal} {\bibinfo
  {journal} {IEEE Trans. Mag.}\ }\textbf {\bibinfo {volume} {49}},\ \bibinfo
  {pages} {4441} (\bibinfo {year} {2013})}\BibitemShut {NoStop}%
\bibitem [{\citenamefont {Lee}\ and\ \citenamefont {Lee}(2016)}]{Lee2016}%
  \BibitemOpen
  \bibfield  {author} {\bibinfo {author} {\bibfnamefont {S.~W.}\ \bibnamefont
  {Lee}}\ and\ \bibinfo {author} {\bibfnamefont {K.~J.}\ \bibnamefont {Lee}},\
  }\href {\doibase 10.1109/JPROC.2016.2543782} {\bibfield  {journal} {\bibinfo
  {journal} {Proc. IEEE}\ }\textbf {\bibinfo {volume} {104}},\ \bibinfo {pages}
  {1831} (\bibinfo {year} {2016})}\BibitemShut {NoStop}%
\bibitem [{\citenamefont {Liu}\ \emph {et~al.}(2012{\natexlab{a}})\citenamefont
  {Liu}, \citenamefont {Pai}, \citenamefont {Li}, \citenamefont {Tseng},
  \citenamefont {Ralph},\ and\ \citenamefont {Buhrman}}]{Liu2012a}%
  \BibitemOpen
  \bibfield  {author} {\bibinfo {author} {\bibfnamefont {L.}~\bibnamefont
  {Liu}}, \bibinfo {author} {\bibfnamefont {C.~F.}\ \bibnamefont {Pai}},
  \bibinfo {author} {\bibfnamefont {Y.}~\bibnamefont {Li}}, \bibinfo {author}
  {\bibfnamefont {H.~W.}\ \bibnamefont {Tseng}}, \bibinfo {author}
  {\bibfnamefont {D.~C.}\ \bibnamefont {Ralph}}, \ and\ \bibinfo {author}
  {\bibfnamefont {R.~A.}\ \bibnamefont {Buhrman}},\ }\href {\doibase
  10.1126/science.1218197} {\bibfield  {journal} {\bibinfo  {journal}
  {Science}\ }\textbf {\bibinfo {volume} {336}},\ \bibinfo {pages} {555}
  (\bibinfo {year} {2012}{\natexlab{a}})}\BibitemShut {NoStop}%
\bibitem [{\citenamefont {Liu}\ \emph {et~al.}(2012{\natexlab{b}})\citenamefont
  {Liu}, \citenamefont {Lee}, \citenamefont {Gudmundsen}, \citenamefont
  {Ralph},\ and\ \citenamefont {Buhrman}}]{Liu2012b}%
  \BibitemOpen
  \bibfield  {author} {\bibinfo {author} {\bibfnamefont {L.}~\bibnamefont
  {Liu}}, \bibinfo {author} {\bibfnamefont {O.~J.}\ \bibnamefont {Lee}},
  \bibinfo {author} {\bibfnamefont {T.~J.}\ \bibnamefont {Gudmundsen}},
  \bibinfo {author} {\bibfnamefont {D.~C.}\ \bibnamefont {Ralph}}, \ and\
  \bibinfo {author} {\bibfnamefont {R.~A.}\ \bibnamefont {Buhrman}},\ }\href
  {\doibase 10.1103/PhysRevLett.109.096602} {\bibfield  {journal} {\bibinfo
  {journal} {Phys. Rev. Lett.}\ }\textbf {\bibinfo {volume} {109}},\ \bibinfo
  {pages} {96602} (\bibinfo {year} {2012}{\natexlab{b}})}\BibitemShut {NoStop}%
\bibitem [{\citenamefont {Miron}\ \emph {et~al.}(2011)\citenamefont {Miron},
  \citenamefont {Garello}, \citenamefont {Gaudin}, \citenamefont {Zermatten},
  \citenamefont {Costache}, \citenamefont {Auffret}, \citenamefont {Bandiera},
  \citenamefont {Rodmacq}, \citenamefont {Schuhl},\ and\ \citenamefont
  {Gambardella}}]{Miron2011}%
  \BibitemOpen
  \bibfield  {author} {\bibinfo {author} {\bibfnamefont {I.~M.}\ \bibnamefont
  {Miron}}, \bibinfo {author} {\bibfnamefont {K.}~\bibnamefont {Garello}},
  \bibinfo {author} {\bibfnamefont {G.}~\bibnamefont {Gaudin}}, \bibinfo
  {author} {\bibfnamefont {P.~J.}\ \bibnamefont {Zermatten}}, \bibinfo {author}
  {\bibfnamefont {M.~V.}\ \bibnamefont {Costache}}, \bibinfo {author}
  {\bibfnamefont {S.}~\bibnamefont {Auffret}}, \bibinfo {author} {\bibfnamefont
  {S.}~\bibnamefont {Bandiera}}, \bibinfo {author} {\bibfnamefont
  {B.}~\bibnamefont {Rodmacq}}, \bibinfo {author} {\bibfnamefont
  {A.}~\bibnamefont {Schuhl}}, \ and\ \bibinfo {author} {\bibfnamefont
  {P.}~\bibnamefont {Gambardella}},\ }\href {\doibase 10.1038/nature10309}
  {\bibfield  {journal} {\bibinfo  {journal} {Nature}\ }\textbf {\bibinfo
  {volume} {476}},\ \bibinfo {pages} {189} (\bibinfo {year}
  {2011})}\BibitemShut {NoStop}%
\bibitem [{\citenamefont {Haney}\ \emph {et~al.}(2013)\citenamefont {Haney},
  \citenamefont {Lee}, \citenamefont {Lee}, \citenamefont {Manchon},\ and\
  \citenamefont {Stiles}}]{Hane2013}%
  \BibitemOpen
  \bibfield  {author} {\bibinfo {author} {\bibfnamefont {P.~M.}\ \bibnamefont
  {Haney}}, \bibinfo {author} {\bibfnamefont {H.~W.}\ \bibnamefont {Lee}},
  \bibinfo {author} {\bibfnamefont {K.~J.}\ \bibnamefont {Lee}}, \bibinfo
  {author} {\bibfnamefont {A.}~\bibnamefont {Manchon}}, \ and\ \bibinfo
  {author} {\bibfnamefont {M.~D.}\ \bibnamefont {Stiles}},\ }\href {\doibase
  10.1103/PhysRevB.87.174411} {\bibfield  {journal} {\bibinfo  {journal} {Phys.
  Rev. B}\ }\textbf {\bibinfo {volume} {87}},\ \bibinfo {pages} {174411}
  (\bibinfo {year} {2013})}\BibitemShut {NoStop}%
\bibitem [{\citenamefont {Berger}(1996)}]{Berger1996}%
  \BibitemOpen
  \bibfield  {author} {\bibinfo {author} {\bibfnamefont {L.}~\bibnamefont
  {Berger}},\ }\href {\doibase 10.1103/PhysRevB.54.9353} {\bibfield  {journal}
  {\bibinfo  {journal} {Phys. Rev. B}\ }\textbf {\bibinfo {volume} {54}},\
  \bibinfo {pages} {9353} (\bibinfo {year} {1996})}\BibitemShut {NoStop}%
\bibitem [{\citenamefont {Slonczewski}(1996)}]{Slonczewski1996}%
  \BibitemOpen
  \bibfield  {author} {\bibinfo {author} {\bibfnamefont {J.~C.}\ \bibnamefont
  {Slonczewski}},\ }\href {\doibase 10.1016/0304-8853(96)00062-5} {\bibfield
  {journal} {\bibinfo  {journal} {J. Mag. Mat.}\ }\textbf {\bibinfo {volume}
  {159}},\ \bibinfo {pages} {L1 } (\bibinfo {year} {1996})}\BibitemShut
  {NoStop}%
\bibitem [{\citenamefont {Ikeda}\ \emph {et~al.}(2010)\citenamefont {Ikeda},
  \citenamefont {Miura}, \citenamefont {Yamamoto}, \citenamefont {Mizunuma},
  \citenamefont {Gan}, \citenamefont {Endo}, \citenamefont {Kanai},
  \citenamefont {Hayakawa}, \citenamefont {Matsukura},\ and\ \citenamefont
  {Ohno}}]{Ikeda2010}%
  \BibitemOpen
  \bibfield  {author} {\bibinfo {author} {\bibfnamefont {S.}~\bibnamefont
  {Ikeda}}, \bibinfo {author} {\bibfnamefont {K.}~\bibnamefont {Miura}},
  \bibinfo {author} {\bibfnamefont {H.}~\bibnamefont {Yamamoto}}, \bibinfo
  {author} {\bibfnamefont {K.}~\bibnamefont {Mizunuma}}, \bibinfo {author}
  {\bibfnamefont {H.~D.}\ \bibnamefont {Gan}}, \bibinfo {author} {\bibfnamefont
  {M.}~\bibnamefont {Endo}}, \bibinfo {author} {\bibfnamefont {S.}~\bibnamefont
  {Kanai}}, \bibinfo {author} {\bibfnamefont {J.}~\bibnamefont {Hayakawa}},
  \bibinfo {author} {\bibfnamefont {F.}~\bibnamefont {Matsukura}}, \ and\
  \bibinfo {author} {\bibfnamefont {H.}~\bibnamefont {Ohno}},\ }\href {\doibase
  10.1038/nmat2804} {\bibfield  {journal} {\bibinfo  {journal} {Nat. Mat.}\
  }\textbf {\bibinfo {volume} {9}},\ \bibinfo {pages} {721} (\bibinfo {year}
  {2010})}\BibitemShut {NoStop}%
\bibitem [{\citenamefont {Yang}\ \emph {et~al.}(2011)\citenamefont {Yang},
  \citenamefont {Chshiev}, \citenamefont {Dieny}, \citenamefont {Lee},
  \citenamefont {Manchon},\ and\ \citenamefont {Shin}}]{Yang2011}%
  \BibitemOpen
  \bibfield  {author} {\bibinfo {author} {\bibfnamefont {H.~X.}\ \bibnamefont
  {Yang}}, \bibinfo {author} {\bibfnamefont {M.}~\bibnamefont {Chshiev}},
  \bibinfo {author} {\bibfnamefont {B.}~\bibnamefont {Dieny}}, \bibinfo
  {author} {\bibfnamefont {J.~H.}\ \bibnamefont {Lee}}, \bibinfo {author}
  {\bibfnamefont {A.}~\bibnamefont {Manchon}}, \ and\ \bibinfo {author}
  {\bibfnamefont {K.~H.}\ \bibnamefont {Shin}},\ }\href {\doibase
  10.1103/PhysRevB.84.054401} {\bibfield  {journal} {\bibinfo  {journal} {Phys.
  Rev. B}\ }\textbf {\bibinfo {volume} {84}},\ \bibinfo {pages} {54401}
  (\bibinfo {year} {2011})}\BibitemShut {NoStop}%
\bibitem [{\citenamefont {Zhao}\ \emph {et~al.}(2015)\citenamefont {Zhao},
  \citenamefont {Jamali}, \citenamefont {Smith},\ and\ \citenamefont
  {Wang}}]{Zhao2015}%
  \BibitemOpen
  \bibfield  {author} {\bibinfo {author} {\bibfnamefont {Z.}~\bibnamefont
  {Zhao}}, \bibinfo {author} {\bibfnamefont {M.}~\bibnamefont {Jamali}},
  \bibinfo {author} {\bibfnamefont {A.~K.}\ \bibnamefont {Smith}}, \ and\
  \bibinfo {author} {\bibfnamefont {J.~P.}\ \bibnamefont {Wang}},\ }\href
  {\doibase 10.1063/1.4916665} {\bibfield  {journal} {\bibinfo  {journal}
  {Appl. Phys. Lett.}\ }\textbf {\bibinfo {volume} {106}},\ \bibinfo {pages}
  {132404} (\bibinfo {year} {2015})}\BibitemShut {NoStop}%
\bibitem [{\citenamefont {Ueda}\ \emph {et~al.}(2016)\citenamefont {Ueda},
  \citenamefont {Mann}, \citenamefont {Pai}, \citenamefont {Tan},\ and\
  \citenamefont {Beach}}]{Ueda2016}%
  \BibitemOpen
  \bibfield  {author} {\bibinfo {author} {\bibfnamefont {K.}~\bibnamefont
  {Ueda}}, \bibinfo {author} {\bibfnamefont {M.}~\bibnamefont {Mann}}, \bibinfo
  {author} {\bibfnamefont {C.~F.}\ \bibnamefont {Pai}}, \bibinfo {author}
  {\bibfnamefont {A.~J.}\ \bibnamefont {Tan}}, \ and\ \bibinfo {author}
  {\bibfnamefont {G.~S.~D.}\ \bibnamefont {Beach}},\ }\href {\doibase
  10.1063/1.4971393} {\bibfield  {journal} {\bibinfo  {journal} {Appl. Phys.
  Lett.}\ }\textbf {\bibinfo {volume} {109}},\ \bibinfo {pages} {232403}
  (\bibinfo {year} {2016})}\BibitemShut {NoStop}%
\bibitem [{\citenamefont {Roschewsky}\ \emph {et~al.}(2016)\citenamefont
  {Roschewsky}, \citenamefont {Matsumura}, \citenamefont {Cheema},
  \citenamefont {Hellman}, \citenamefont {Kato}, \citenamefont {Iwata},\ and\
  \citenamefont {Salahuddin}}]{Roschewsky2016}%
  \BibitemOpen
  \bibfield  {author} {\bibinfo {author} {\bibfnamefont {N.}~\bibnamefont
  {Roschewsky}}, \bibinfo {author} {\bibfnamefont {T.}~\bibnamefont
  {Matsumura}}, \bibinfo {author} {\bibfnamefont {S.}~\bibnamefont {Cheema}},
  \bibinfo {author} {\bibfnamefont {F.}~\bibnamefont {Hellman}}, \bibinfo
  {author} {\bibfnamefont {T.}~\bibnamefont {Kato}}, \bibinfo {author}
  {\bibfnamefont {S.}~\bibnamefont {Iwata}}, \ and\ \bibinfo {author}
  {\bibfnamefont {S.}~\bibnamefont {Salahuddin}},\ }\href {\doibase
  10.1063/1.4962812} {\bibfield  {journal} {\bibinfo  {journal} {Appl. Phys.
  Lett.}\ }\textbf {\bibinfo {volume} {109}},\ \bibinfo {pages} {112403}
  (\bibinfo {year} {2016})}\BibitemShut {NoStop}%
\bibitem [{\citenamefont {Finley}\ and\ \citenamefont
  {Liu}(2016)}]{Finley2016}%
  \BibitemOpen
  \bibfield  {author} {\bibinfo {author} {\bibfnamefont {J.}~\bibnamefont
  {Finley}}\ and\ \bibinfo {author} {\bibfnamefont {L.}~\bibnamefont {Liu}},\
  }\href {\doibase 10.1103/PhysRevApplied.6.054001} {\bibfield  {journal}
  {\bibinfo  {journal} {Phys. Rev. Applied}\ }\textbf {\bibinfo {volume} {6}},\
  \bibinfo {pages} {54001} (\bibinfo {year} {2016})}\BibitemShut {NoStop}%
\bibitem [{\citenamefont {Fan}\ \emph {et~al.}(2014)\citenamefont {Fan},
  \citenamefont {Celik}, \citenamefont {Wu}, \citenamefont {Ni}, \citenamefont
  {Lee}, \citenamefont {Lorenz},\ and\ \citenamefont {Xiao}}]{Fan2014}%
  \BibitemOpen
  \bibfield  {author} {\bibinfo {author} {\bibfnamefont {X.}~\bibnamefont
  {Fan}}, \bibinfo {author} {\bibfnamefont {H.}~\bibnamefont {Celik}}, \bibinfo
  {author} {\bibfnamefont {J.}~\bibnamefont {Wu}}, \bibinfo {author}
  {\bibfnamefont {C.}~\bibnamefont {Ni}}, \bibinfo {author} {\bibfnamefont
  {K.~J.}\ \bibnamefont {Lee}}, \bibinfo {author} {\bibfnamefont {V.~O.}\
  \bibnamefont {Lorenz}}, \ and\ \bibinfo {author} {\bibfnamefont {J.~Q.}\
  \bibnamefont {Xiao}},\ }\href {\doibase 10.1038/ncomms4042} {\bibfield
  {journal} {\bibinfo  {journal} {Nat. Comm.}\ }\textbf {\bibinfo {volume}
  {5}},\ \bibinfo {pages} {3042} (\bibinfo {year} {2014})}\BibitemShut
  {NoStop}%
\bibitem [{\citenamefont {{Lo Conte}}\ \emph {et~al.}(2016)\citenamefont {{Lo
  Conte}}, \citenamefont {Karnad}, \citenamefont {Martinez}, \citenamefont
  {Lee}, \citenamefont {Kim}, \citenamefont {Han}, \citenamefont {Kim},
  \citenamefont {Prenzel}, \citenamefont {Schulz}, \citenamefont {You},
  \citenamefont {Swagten},\ and\ \citenamefont {Klaeui}}]{Conte2016}%
  \BibitemOpen
  \bibfield  {author} {\bibinfo {author} {\bibfnamefont {R.}~\bibnamefont {{Lo
  Conte}}}, \bibinfo {author} {\bibfnamefont {G.~V.}\ \bibnamefont {Karnad}},
  \bibinfo {author} {\bibfnamefont {E.}~\bibnamefont {Martinez}}, \bibinfo
  {author} {\bibfnamefont {K.}~\bibnamefont {Lee}}, \bibinfo {author}
  {\bibfnamefont {N.~H.}\ \bibnamefont {Kim}}, \bibinfo {author} {\bibfnamefont
  {D.~S.}\ \bibnamefont {Han}}, \bibinfo {author} {\bibfnamefont {J.~S.}\
  \bibnamefont {Kim}}, \bibinfo {author} {\bibfnamefont {S.}~\bibnamefont
  {Prenzel}}, \bibinfo {author} {\bibfnamefont {T.}~\bibnamefont {Schulz}},
  \bibinfo {author} {\bibfnamefont {C.~Y.}\ \bibnamefont {You}}, \bibinfo
  {author} {\bibfnamefont {H.~J.~M.}\ \bibnamefont {Swagten}}, \ and\ \bibinfo
  {author} {\bibfnamefont {M.}~\bibnamefont {Klaeui}},\ }\href@noop {}
  {\bibfield  {journal} {\bibinfo  {journal} {ArXiv}\ } (\bibinfo {year}
  {2016})},\ \Eprint {http://arxiv.org/abs/1609.02078} {arXiv:1609.02078}
  \BibitemShut {NoStop}%
\bibitem [{\citenamefont {Lee}\ \emph {et~al.}(2014)\citenamefont {Lee},
  \citenamefont {Lee}, \citenamefont {Cho}, \citenamefont {Choi}, \citenamefont
  {You}, \citenamefont {Jung}, \citenamefont {Bonell}, \citenamefont {Shiota},
  \citenamefont {Miwa},\ and\ \citenamefont {Suzuki}}]{Lee2014}%
  \BibitemOpen
  \bibfield  {author} {\bibinfo {author} {\bibfnamefont {H.~R.}\ \bibnamefont
  {Lee}}, \bibinfo {author} {\bibfnamefont {K.}~\bibnamefont {Lee}}, \bibinfo
  {author} {\bibfnamefont {J.}~\bibnamefont {Cho}}, \bibinfo {author}
  {\bibfnamefont {Y.~H.}\ \bibnamefont {Choi}}, \bibinfo {author}
  {\bibfnamefont {C.~Y.}\ \bibnamefont {You}}, \bibinfo {author} {\bibfnamefont
  {M.~H.}\ \bibnamefont {Jung}}, \bibinfo {author} {\bibfnamefont
  {F.}~\bibnamefont {Bonell}}, \bibinfo {author} {\bibfnamefont
  {Y.}~\bibnamefont {Shiota}}, \bibinfo {author} {\bibfnamefont
  {S.}~\bibnamefont {Miwa}}, \ and\ \bibinfo {author} {\bibfnamefont
  {Y.}~\bibnamefont {Suzuki}},\ }\href {\doibase 10.1038/srep06548} {\bibfield
  {journal} {\bibinfo  {journal} {Sci. Rep.}\ }\textbf {\bibinfo {volume}
  {4}},\ \bibinfo {pages} {6548} (\bibinfo {year} {2014})}\BibitemShut
  {NoStop}%
\bibitem [{\citenamefont {Shirakawa}\ \emph {et~al.}(1976)\citenamefont
  {Shirakawa}, \citenamefont {Nakajima}, \citenamefont {Okamoto}, \citenamefont
  {Matsushita},\ and\ \citenamefont {Sakurai}}]{Shirakawa1976}%
  \BibitemOpen
  \bibfield  {author} {\bibinfo {author} {\bibfnamefont {T.}~\bibnamefont
  {Shirakawa}}, \bibinfo {author} {\bibfnamefont {Y.}~\bibnamefont {Nakajima}},
  \bibinfo {author} {\bibfnamefont {K.}~\bibnamefont {Okamoto}}, \bibinfo
  {author} {\bibfnamefont {S.}~\bibnamefont {Matsushita}}, \ and\ \bibinfo
  {author} {\bibfnamefont {Y.}~\bibnamefont {Sakurai}},\ }\href {\doibase
  10.1063/1.2946125} {\bibfield  {journal} {\bibinfo  {journal} {AIP Conference
  Proceedings}\ }\textbf {\bibinfo {volume} {34}},\ \bibinfo {pages} {349}
  (\bibinfo {year} {1976})}\BibitemShut {NoStop}%
\bibitem [{\citenamefont {Mimura}\ \emph {et~al.}(1976)\citenamefont {Mimura},
  \citenamefont {Imamura},\ and\ \citenamefont {Kushiro}}]{Mimura1976}%
  \BibitemOpen
  \bibfield  {author} {\bibinfo {author} {\bibfnamefont {Y.}~\bibnamefont
  {Mimura}}, \bibinfo {author} {\bibfnamefont {N.}~\bibnamefont {Imamura}}, \
  and\ \bibinfo {author} {\bibfnamefont {Y.}~\bibnamefont {Kushiro}},\ }\href
  {\doibase 10.1063/1.323098} {\bibfield  {journal} {\bibinfo  {journal} {J.
  Appl. Phys.}\ }\textbf {\bibinfo {volume} {47}},\ \bibinfo {pages} {3371}
  (\bibinfo {year} {1976})}\BibitemShut {NoStop}%
\bibitem [{\citenamefont {Hayashi}\ \emph {et~al.}(2014)\citenamefont
  {Hayashi}, \citenamefont {Kim}, \citenamefont {Yamanouchi},\ and\
  \citenamefont {Ohno}}]{Hayashi2014}%
  \BibitemOpen
  \bibfield  {author} {\bibinfo {author} {\bibfnamefont {M.}~\bibnamefont
  {Hayashi}}, \bibinfo {author} {\bibfnamefont {J.}~\bibnamefont {Kim}},
  \bibinfo {author} {\bibfnamefont {M.}~\bibnamefont {Yamanouchi}}, \ and\
  \bibinfo {author} {\bibfnamefont {H.}~\bibnamefont {Ohno}},\ }\href {\doibase
  10.1103/PhysRevB.89.144425} {\bibfield  {journal} {\bibinfo  {journal} {Phys.
  Rev. B}\ }\textbf {\bibinfo {volume} {89}},\ \bibinfo {pages} {144425}
  (\bibinfo {year} {2014})}\BibitemShut {NoStop}%
\bibitem [{\citenamefont {Woo}\ \emph {et~al.}(2014)\citenamefont {Woo},
  \citenamefont {Mann}, \citenamefont {Tan}, \citenamefont {Caretta},\ and\
  \citenamefont {Beach}}]{Woo2014}%
  \BibitemOpen
  \bibfield  {author} {\bibinfo {author} {\bibfnamefont {S.}~\bibnamefont
  {Woo}}, \bibinfo {author} {\bibfnamefont {M.}~\bibnamefont {Mann}}, \bibinfo
  {author} {\bibfnamefont {A.~J.}\ \bibnamefont {Tan}}, \bibinfo {author}
  {\bibfnamefont {L.}~\bibnamefont {Caretta}}, \ and\ \bibinfo {author}
  {\bibfnamefont {G.~S.~D.}\ \bibnamefont {Beach}},\ }\href {\doibase
  10.1063/1.4902529} {\bibfield  {journal} {\bibinfo  {journal} {Appl. Phys.
  Lett.}\ }\textbf {\bibinfo {volume} {105}},\ \bibinfo {pages} {212404}
  (\bibinfo {year} {2014})}\BibitemShut {NoStop}%
\bibitem [{\citenamefont {Lee}\ \emph {et~al.}(2013)\citenamefont {Lee},
  \citenamefont {Lee}, \citenamefont {Min},\ and\ \citenamefont
  {Lee}}]{Lee2013}%
  \BibitemOpen
  \bibfield  {author} {\bibinfo {author} {\bibfnamefont {K.~S.}\ \bibnamefont
  {Lee}}, \bibinfo {author} {\bibfnamefont {S.~W.}\ \bibnamefont {Lee}},
  \bibinfo {author} {\bibfnamefont {B.~C.}\ \bibnamefont {Min}}, \ and\
  \bibinfo {author} {\bibfnamefont {K.~J.}\ \bibnamefont {Lee}},\ }\href
  {\doibase 10.1063/1.4798288} {\bibfield  {journal} {\bibinfo  {journal}
  {Appl. Phys. Lett.}\ }\textbf {\bibinfo {volume} {102}},\ \bibinfo {pages}
  {112410} (\bibinfo {year} {2013})}\BibitemShut {NoStop}%
\bibitem [{\citenamefont {El-Hilo}\ \emph {et~al.}(1992)\citenamefont
  {El-Hilo}, \citenamefont {de~Witte}, \citenamefont {O'Grady},\ and\
  \citenamefont {Chantrell}}]{ElHilo1992}%
  \BibitemOpen
  \bibfield  {author} {\bibinfo {author} {\bibfnamefont {M.}~\bibnamefont
  {El-Hilo}}, \bibinfo {author} {\bibfnamefont {A.~M.}\ \bibnamefont
  {de~Witte}}, \bibinfo {author} {\bibfnamefont {K.}~\bibnamefont {O'Grady}}, \
  and\ \bibinfo {author} {\bibfnamefont {R.~W.}\ \bibnamefont {Chantrell}},\
  }\href {\doibase 10.1016/0304-8853(92)90085-3} {\bibfield  {journal}
  {\bibinfo  {journal} {J. Mag. Mat.}\ }\textbf {\bibinfo {volume} {117}},\
  \bibinfo {pages} {L307 } (\bibinfo {year} {1992})}\BibitemShut {NoStop}%
\end{thebibliography}%

\end{document}